\documentclass[a4paper,fleqn]{cas-dc}

\usepackage[authoryear,longnamesfirst]{natbib}

\usepackage{cite}
\usepackage{subfigure}
\usepackage{tabularx}
\usepackage{float}
\usepackage{amsmath} 
\usepackage{amssymb}
\usepackage{booktabs}
\usepackage{todonotes}
\usepackage{color}
\usepackage{overpic}
\usepackage{subfigure}
\usepackage{bm}
\usepackage{bbding}
\usepackage{makecell}
\usepackage{array}

\usepackage{url}
\usepackage{xcolor}

\usepackage{hyperref}

\begin{document}



\title [mode = title]{End-to-end Triple-domain PET Enhancement: A Hybrid Denoising-and-reconstruction Framework for Reconstructing Standard-dose PET Images from Low-dose PET Sinograms}  



%


\author[1]{{Caiwen~Jiang }}
\author[4]{{Mianxin~Liu}}
\author[1]{{Kaicong~Sun}}
\author[1,2,3]{{Dinggang Shen \corref{cor1}}}
\cortext[cor1]{Corresponding author.}
\ead{dgshen@shanghaitech.edu.cn}

\address[1]{School of Biomedical Engineering \& State Key Laboratory of Advanced Medical Materials and Devices, ShanghaiTech University, Shanghai, China}
\address[2]{Shanghai United Imaging Intelligence Co., Ltd., Shanghai, China}
\address[3]{Shanghai Clinical Research and Trial Center, Shanghai, 201210, China}
\address[4]{Shanghai Artificial Intelligence Laboratory, Shanghai 200232, China}















\begin{abstract}
As a sensitive functional imaging technique, positron emission tomography (PET) plays a critical role in early disease diagnosis. However, obtaining a high-quality PET image requires injecting a sufficient dose (standard dose) of radionuclides into the body, which inevitably poses radiation hazards to patients. To mitigate radiation hazards, the reconstruction of standard-dose PET (SPET) from low-dose PET (LPET) is desired. According to imaging theory, PET reconstruction process involves multiple domains (e.g., projection domain and image domain), and a significant portion of the difference between SPET and LPET arises from variations in the noise levels introduced during the sampling of raw data as sinograms. In light of these two facts, we propose an end-to-end TriPle-domain LPET EnhancemenT (TriPLET) framework, by leveraging the advantages of a hybrid denoising-and-reconstruction process and a triple-domain representation (i.e., sinograms, frequency spectrum maps, and images) to reconstruct SPET images from LPET sinograms. Specifically, TriPLET consists of three sequentially coupled components including 1) a Transformer-assisted denoising network that denoises the inputted LPET sinograms in the projection domain, 2) a discrete-wavelet-transform-based reconstruction network that further reconstructs SPET from LPET in the wavelet domain, and 3) a pair-based adversarial network that evaluates the reconstructed SPET images in the image domain. Extensive experiments on the real PET dataset demonstrate that our proposed TriPLET can reconstruct SPET images with the highest similarity and signal-to-noise ratio to real data, compared with state-of-the-art methods. 
\end{abstract}




\maketitle

\section{Introduction}

\begin{figure*}[!t]
 \setlength{\abovecaptionskip}{0.1cm}
\setlength{\belowcaptionskip}{-0.4cm}
\centering
\begin{overpic}[width=1\linewidth]{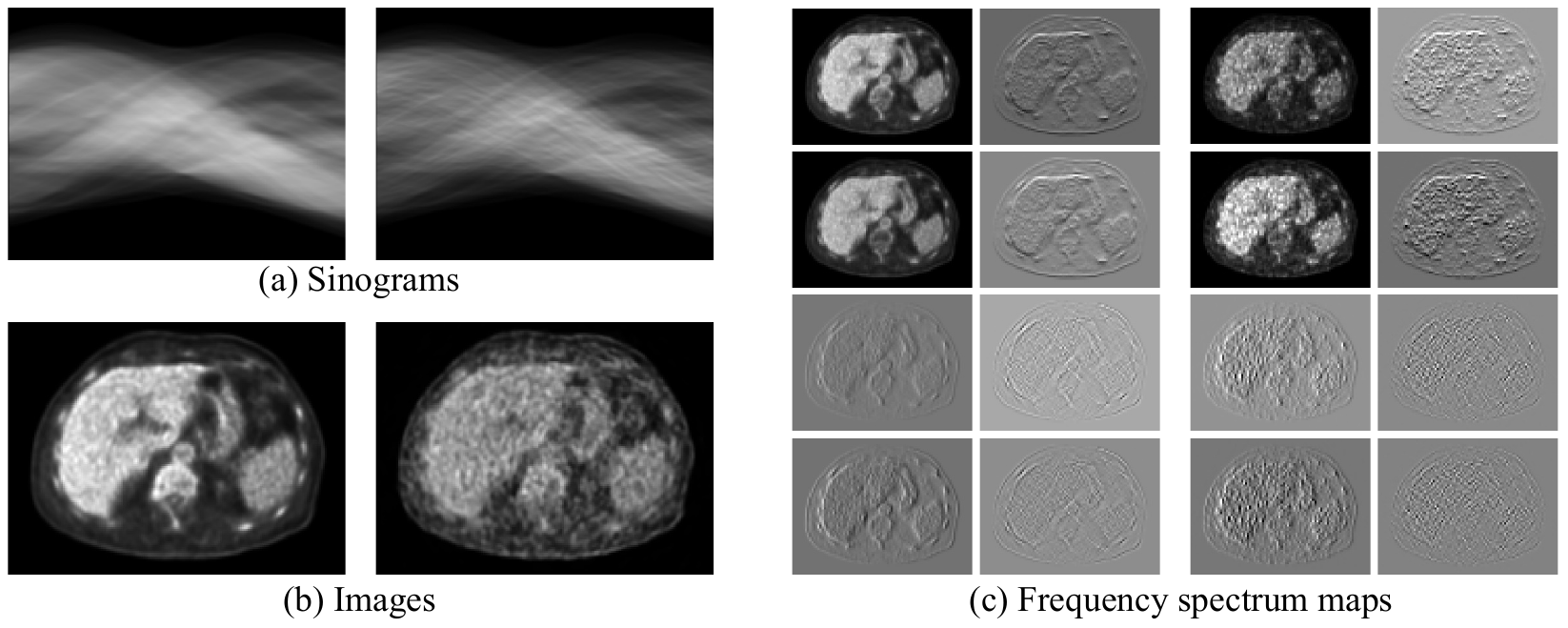}

    \end{overpic}
    \vspace{-4mm}
\centering
\caption{Representation of PET data in three domains: (a) projection domain, (b) image domain, and (c) wavelet domain, with the SPET and LPET shown on the left and right of each subfigure, respectively. Radon transform and discrete wavelet transform can convert PET data between images and sinograms, and also between images and frequency spectrum maps, respectively. After applying discrete wavelet transform, a 3D PET image can be decomposed into eight frequency spectrum maps of different sub-bands.}
\label{case}
\end{figure*}
Positron emission tomography (PET) is a non-invasive and sensitive imaging technique with an essential role in early disease diagnosis and intervention in clinics~\citep{2020Noise,chen2007clinical,luo20213d,sakthivel2020clinical}. By injecting the radionuclides into the body and capturing the emitted positron signals, PET can visualize the metabolic and biochemical processes in the body by assessing the radionuclides' distributions~\citep{2005Positron,decazes2021trimodality}. This characteristic makes PET a valuable imaging modality for precise diagnoses of diseases, particularly for diseases such as cancer and Alzheimer's disease ~\citep{jiang2022reconstruction,dimitrakopoulou2021kinetic}.

But the radionuclides used in PET imaging are radioactive, which inevitably poses a risk of radiation exposure to patients~\citep{2012Oncologic}. In addition,  due to limitations in signal receiver sensitivity and noise interference, the dose of radionuclides must meet a certain threshold to produce PET images with sufficient quality for clinical diagnosis~\citep{2021Standard,kreisl2020pet}. Despite applying the As Low As Reasonably Achievable (ALARA)~\citep{Slovis2002The} principle in clinical imaging to minimize radiation exposure, PET imaging may still be deemed unacceptable for certain populations, such as pediatric subjects and pregnant women~\citep{2020Estimating,de2020nonprostatic}. To mitigate the radiation risks associated with PET imaging, designing advanced reconstruction algorithms to enhance PET image quality (e.g., reconstructing standard-dose PET (SPET) from low-dose PET (LPET)) is a promising alternative.

In recent decades, many PET enhancement methods have been developed, with most of those methods focusing only on SPET reconstruction in a single domain, typically the image domain. However, it is important to note that during the reconstruction process, PET data are not limited to images alone but can be represented in multiple forms, each emphasizing different aspects of information. For instance, PET data can be represented as sinograms, images, and frequency spectrum maps in the projection, image, and wavelet domains, respectively, as depicted in Figure~\ref{case}. These domains can be converted to each other in a lossless fashion using the Radon transform~\citep{hamill2003fast} and discrete wavelet transform (WT)~\citep{bhavana2015multi}. Therefore, the incorporation of multiple domains in the design of reconstruction algorithms is expected to lead to significant improvements. 

In addition, another important fact is that PET data is initially sampled as a sinogram and then subsequently reconstructed into an image~\citep{fahey2002data}. Furthermore, a significant distinction between LPET and SPET lies in the discrepancy in noise levels within the sinogram~\citep{mokri2016pet}. Thus, it is useful to adopt a hybrid denoising-and-reconstruction framework for incorporating a sinogram noise reduction step before reconstruction. This approach can align more closely with the actual reconstruction process and effectively suppress the propagation and amplification of noises inherent in the raw data (sinogram) during the subsequent processing.

Accordingly, we propose an end-to-end TriPle-domain LPET EnhancemenT (TriPLET) framework, by considering a hybrid denoising-and-reconstruction process and simultaneously taking the advantages of projection, wavelet, and image domains in reconstructing SPET from LPET. The TriPLET is based on our previous work TriDoRNet~\citep{jiang2023tridornet} and comprises 1) a denoising network to initially remove noises from the inputted LPET sinograms in the projection domain, 2) a reconstruction network to refine the denoising process and also reconstruct SPET from LPET in the wavelet domain, and 3) an adversarial network to evaluate the reconstructed SPET images in the image domain. The most significant differences between TriPLET and TriDoRNet include 1) The denoising network architecture has been redesigned, i.e., in the current work, the denoising network adopts an alternating structure of CNN and transformer blocks to achieve improved denoising effects. 2) An adversarial network has been integrated into the framework, and a paired adversarial loss has been designed to make the reconstructed SPET images more closely resemble real images. 3) The GradNorm~\citep{chen2018gradnorm} technique has been used to automatically adjust the weights of different loss functions during training, reducing the difficulty of tuning parameters. 4) Due to the use of different methodologies, the related research background, literature review, and experimental design have also been different from our previous work.

Among these networks, the denoising network adopts a residual learning strategy to predict the residual sinogram, and further integrates Transformer modules to capture noises as long-range representations contained in the sinogram. The reconstruction network follows the 3D U-Net architecture, but replaces both downsampling and upsampling operations with wavelet transform (WT) and inverse wavelet transform (IWT), respectively, to allow the network to work on the wavelet frequency domains for preserving high-frequency structural details. The adversarial network takes two pairs of images, i.e., the predicted SPET $\&$ LPET images and the real SPET $\&$ LPET images, as the inputs, to properly discriminate the fake image pair from the real image pair for facilitating accurate reconstruction. Furthermore, to improve the supervision in training the TriPLET, we design an integrated loss function by combining constraints in the projection, wavelet, and image domains.

The main contributions of this work can be summarized as follows:\\
\begin{enumerate}
    \item 
To leverage the advantages of projection, wavelet, and image domains in reconstructing SPET from LPET, we propose an end-to-end TriPle-domain LPET EnhancemenT (TriPLET) framework coupled by a denoising network, a reconstruction network, and an adversarial network. 
    \item 
To facilitate the reconstruction, we adopt a hybrid denoising-and-reconstruction framework by incorporating a sinogram noise reduction step before reconstruction. This framework allows an end-to-end reconstruction in the practical sampling process. Additionally, we design a transformer-based denoising network to capture the long-range information in the LPET sinogram.

    \item 
To enhance the structural details of the reconstructed SPET images, we incorporate discrete wavelet transform into the reconstruction network, which enables the network to capture important frequency band information of PET.

    \item
 Extensive experiments conducted on clinical chest-abdomen PET data show superior performance of our approach over state-of-the-art methods in SPET image reconstruction. 
\end{enumerate}

The subsequent sections of this paper are structured as follows. Section \uppercase\expandafter{\romannumeral2} provides an overview of the related work, while Section \uppercase\expandafter{\romannumeral3} presents a detailed description of our proposed approach. Implementation details and experimental results are presented in Section \uppercase\expandafter{\romannumeral4}. Finally, our work is summarized in Section \uppercase\expandafter{\romannumeral5}.

\section{Related Work}
In this section, we provide a comprehensive review of related works in two specific domains, including PET enhancement and multi-domain image reconstruction, to better illustrate our SPET enhancement task and our motivation for leveraging multi-domain representation into the reconstruction process.

\subsection{PET Enhancement}
In recent years, many enhancement algorithms have been developed to improve PET image quality. The first set of methods are the filtering-based methods that are used to remove noise from LPET images. Typical methods include the non-local mean (NLM) filter with or without anatomical information~\citep{buades2005non, chan2009non}, block-matching 3D (BM3D) filtering and its higher dimensional variants BM4D and BM5D~\citep{dabov2006image,ote2020kinetics}, wavelet filter (for removing noise from specific frequency sub-bands)~\citep{le2013denoising}, and guided ﬁlter (by integrating cross-modality information)~\citep{yan2015mri}. These methods are quite robust, but tend to over-smooth images and suppress high-frequency details.

The second set of methods is the machine learning-based algorithms.  For instance, Kang \textsl{et al}.~\citep{2015Prediction} introduced a random forest-based approach to predict SPET images at the voxel level. Wang \textsl{et al}.~\citep{2015Predicting} proposed a mapping-based sparse learning method that leverages LPET images and corresponding MR images for SPET image prediction. An \textsl{et al}.~\citep{2016Multi} developed a data-driven multi-level correlation analysis methodology that utilizes LPET and MRI images for SPET image generation. However, due to the unavailability of corresponding MR images for most PET images, Wang \textsl{et al}.~\citep{2016Semisupervised} proposed a semi-supervised triple dictionary learning method that incorporates a large number of unpaired training samples in the SPET image generation process. Nonetheless, these methods often require manual intervention and are time-consuming, which hinders their widespread application in clinic.

Currently, deep learning-based methods, leveraging the advantages of convolutional neural networks (CNNs) in image processing~\citep{2016Deep,2017BrainNetCNN}, have emerged as the dominant approach for PET enhancement. Several mapping networks, such as 3D CNN~\citep{2018Penalized}, conditional GAN~\citep{20183D}, locally adaptive GAN~\citep{20193D}, and adaptive rectiﬁcation GAN~\citep{luo2022adaptive}, have been proposed to generate SPET images from LPET images. However, a common drawback of these methods is their reliance on paired LPET and SPET images for supervision. To overcome this limitation, unsupervised methods based on deep image prior and Noise2Noise have been introduced. For instance, Cui \textsl{et al}.~\citep{cui2019pet} proposed an unsupervised deep learning-based method for PET denoising, which utilizes a prior high-quality image from the patient as the network input and the noisy PET image itself as the training label. Onishi \textsl{et al}.~\citep{onishi2021anatomical} developed an unsupervised PET image denoising method that incorporates an anatomical information-guided attention mechanism, leveraging anatomical information from MR to guide PET denoising. However, these unsupervised methods require alternative-modality images, such as MR/CT, to serve as guiding images, and the predicted PET images often retain some residual content from MR/CT due to the lack of explicit supervision. Additionally, some studies have explored the direct reconstruction of SPET images from LPET sinograms. For example, Häggström \textsl{et al}.~\citep{haggstrom2019deeppet} proposed a deep learning model, called DeepPET, which reconstructs PET images directly from sinogram data without relying on system and noise models. Feng \textsl{et al}.~\citep{feng2020rethinking} introduced a PET reconstruction framework that employs two coupled networks to directly reconstruct high-quality PET images from ultra-low-dose sinograms. These deep learning-based methods are significantly different from time-consuming traditional PET reconstruction algorithms, such as Ordered Subset Expectation Maximization (OSEM)~\citep{byrd2023impact}, since, in the application stage, the reconstructed image can be obtained by going through a single forward the trained network (without iterative optimization).

\begin{figure*}[!t]
\centering
\begin{overpic}[width=1\linewidth]{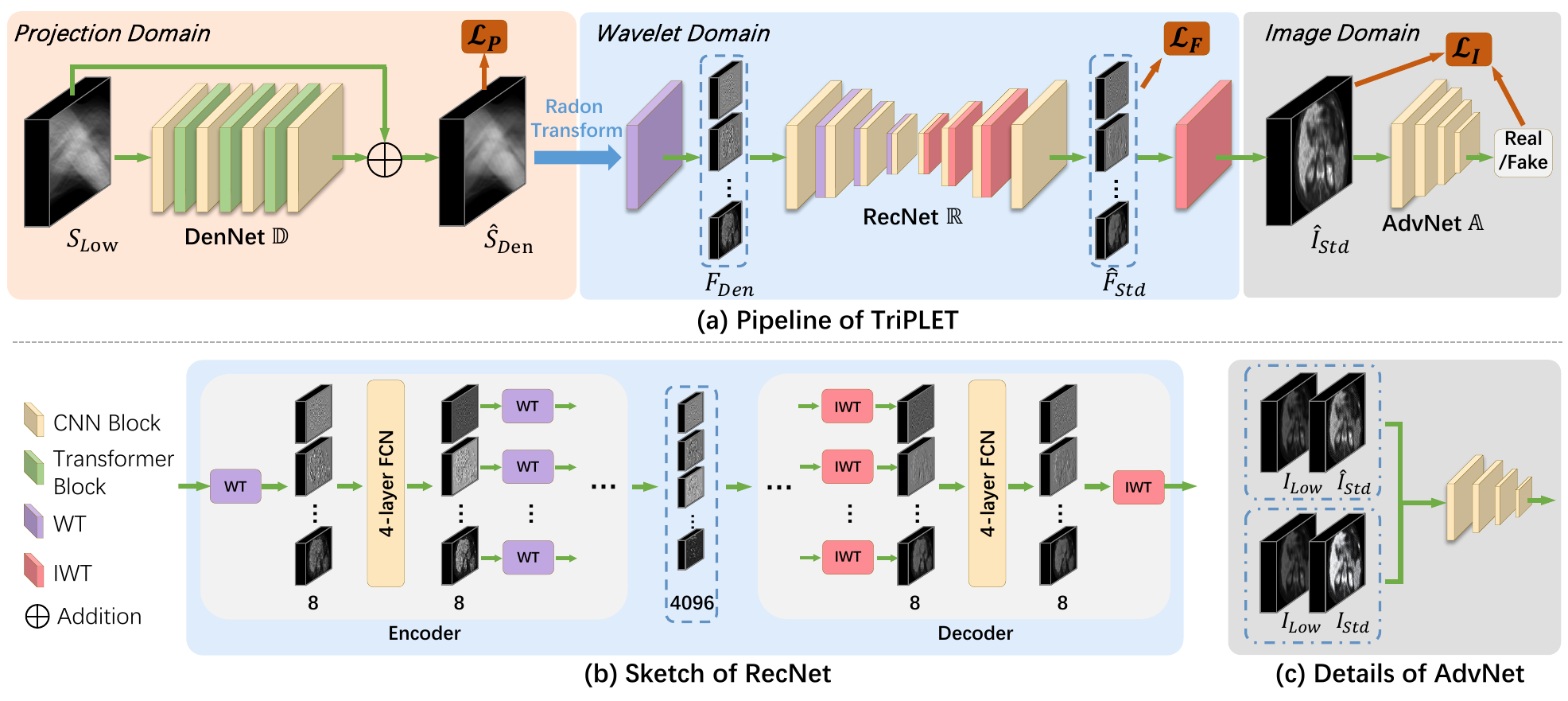}

    \end{overpic}
    \vspace{-5mm}
\centering 
\caption{Our proposed TriPLET involves three domains, i.e., the projection, wavelet, and image domains. PET data have different representations in each domain, such as the LPET sinogram $S_{Low}$ and the denoised sinogram $S_{Den}$ in the projection domain, the denoised frequency spectrum map $F_{Den}$ and the predicted SPET map $\hat{F}_{Std}$ in the wavelet domain, and the LPET image $I_{Low}$, the predicted SPET image $\hat{I}_{Std}$, and the actual SPET image $I_{Std}$ in the image domain. We have also designed specific loss functions  $\mathcal{L}_{P}$, $\mathcal{L}_{F}$, and $\mathcal{L}_{I}$ to supervise the prediction results in each domain.}
\label{fig_framework}
\end{figure*}

\subsection{Multi-domain Image Reconstruction}
According to imaging theory, multiple domains are involved in reconstructing medical images (e.g., PET, MR, and CT) from the original signals, thus considering multiple domains in designing reconstruction algorithms may achieve successful improvement. This has been demonstrated in various studies. For instance, Arabi \textsl{et al}.~\citep{arabi2018improvement} introduced a hybrid dual-domain PET denoising method that combines advantages of both image and projection domains to preserve image textures while minimizing quantification uncertainty. Souza \textsl{et al}.~\citep{souza2019hybrid} proposed a hybrid frequency-/image-domain cascade of convolutional neural networks (CNNs) with intercalated data consistency layers for MR image reconstruction. Wu \textsl{et al}.~\citep{wu2021drone} developed a dual-domain residual-based optimization network for CT image reconstruction, sequentially operating in the projection and image domains. Zhou \textsl{et al}.~\citep{zhou2022dudodr} presented a dual-domain data consistent recurrent network that reconstructs artifact-free images through recurrent image domain and projection domain restorations. The successes of these previous works have served as inspiration for our design of the end-to-end triple-domain PET enhancement (TriPLET) framework.

\subsection{Adversarial Learning in Medical Image}
Adversarial learning~\citep{goodfellow2014generative} is an essential technique in the deep learning field. It involves training by placing a generative model and a discriminative model in an adversarial framework, where the generative model aims to generate data as close to real data as possible, and the discriminative model tries to distinguish between real and generated data. This setup forces both generative and discriminative models to continuously improve during the training process, thereby enhancing the quality of the generated data. Due to this advantage, adversarial learning strategies have been widely adopted in many medical image analysis studies~\citep{yi2019generative}. For example, Hu \textsl{et al}.~\citep{hu2020medical} develop a data augmentation method based on generative adversarial networks to reconstruct missing PET images from MRI for AD assessment. Sun \textsl{et al}.~\citep{sun2020adversarial} propose a medical image synthesis model named abnormal-to-normal translation generative adversarial network (ANT-GAN) to generate a normal-looking medical image based on its abnormal-looking counterpart without the need for paired training data. Li \textsl{et al}.~\citep{li2021novel} introduce a medical image denoising method based on conditional Generative Adversarial Networks (cGAN) for medical image denoising tasks.

\section{Method}
Figure~\ref{fig_framework} (a) illustrates the pipeline of our TriPLET. When given an LPET sinogram $S_{Low}$, TriPLET first applies the denoising network (DenNet) $\mathbb{D}$ to denoise it in the projection domain. Next, the reconstruction network (RecNet) $\mathbb{R}$ reconstructs the SPET image $I_{Std}$ from the denoised sinogram $S_{Den}$ in the wavelet domain. Finally, the adversarial network (AdvNet) $\mathbb{A}$ evaluates the predicted SPET image $\hat{I}_{Std}$ in the image domain. Moreover, we design respective loss functions, including $\mathcal{L}_{P}$, $\mathcal{L}_{F}$, and $\mathcal{L}_{I}$, to supervise the prediction results of each network. Subsequently, we introduce the construction of sub-networks and the design details of each loss function.

\subsection{Denoising Network}
Compared to SPET, LPET exhibits stronger noise during imaging due to its weaker signal strength, which is directly reflected in the raw data of PET, i.e., the sinogram. To effectively prevent the propagation and amplification of noise during the subsequent processing and also align our enhancement framework more closely with the actual PET reconstruction process, we incorporate a noise reduction procedure for sinogram before reconstruction. Hence, we design a denoising network (DenNet) $\mathbb{D}$ in the projection domain to denoise the LPET sinogram $S_{Low}$. As shown in Figure~\ref{fig_framework} (a), $\mathbb{D}$ uses the residual learning strategy to predict residual sinograms with seven blocks, where the \{$1^{st}$, $3^{rd}$, $5^{th}$, $7^{th}$\} and \{$2^{nd}$, $4^{th}$, $6^{th}$\} blocks are the CNN blocks and Transformer blocks, respectively. Its uniqueness lies in the \textit{Residual Learning Strategy} and \textit{Capturing Long-range Information via Transformer}, as described below.
\citep{song2021noise2void}

\textsl{Residual Learning Strategy:}
We design the denoising network $\mathbb{D}$ based on the concept of residual learning~\citep{zhang2017beyond}, which is commonly utilized in various denoising tasks. Specifically, $\mathbb{D}$ is designed to predict the residual sinogram which represents the difference between denoised sinogram $S_{Den}$ and SPET sinogram $S_{Std}$ rather than directly predicting the $S_{Den}$. In this way, we can estimate noisy components (the residual parts) and effectively avoid gradient vanishing while also enhancing the network's ability to learn the underlying features of the sinogram by enabling it to have a larger number of layers.

\textsl{Capturing Long-range Information via Transformer:} 
According to projection theory, the signals collected in PET imaging are represented as the line of response (LOR) in the sinogram. This means that noise is also represented as LOR in the sinogram with non-local (long-range) representation. The limited perceptual field in the conventional convolutional neural network makes it hard to capture such long-range information. Therefore, we incorporate CNNs with the Transformer~\citep{ liu2021swin} blocks into the denoising network to address this issue. Compared to CNN, the Transformer is better at capturing long-range information, making it more suitable for the sinogram with long-range representation but at the cost of higher computational demands. Considering the balance between performance and computational cost, our denoising network employs a hybrid structure of CNN and Transformer blocks. Moreover, alternatingly using CNN and Transformer blocks allows for the effective extraction of local information by CNN blocks and global information by Transformer blocks, which has been successful in some studies~\citep{geng2024stcnet,yang2023alternating}. Inspired by the success of these works, our denoising network ultimately adopts an alternating combination of 3 CNN blocks and 3 Transformer blocks.

\begin{figure}[!t]
\centering
\begin{overpic}[width=1\linewidth]{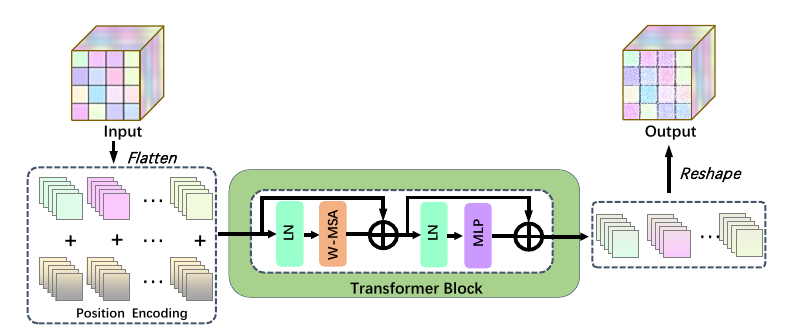}

    \end{overpic}

\centering 
\caption{Architectural detail of the Transformer block in the DenNet (denoising network).}
\label{fig_transformer}
\end{figure}

For the CNN blocks, we employ a Conv+BN+ReLU architecture. Specifically, Conv is the convolutional layer with a kernel size of 3×3×3 and a stride of 1, BN is the batch normalization with a batch size of 4, and ReLU stands for rectified linear units, which are used for introducing non-linearity. The features outputted by the CNN block are subsequently inputted into the Transformer block. Figure~\ref{fig_transformer} illustrates the architectural details of the Transformer blocks. Each Transformer block consists of a window-based multi-head self-attention (W-MSA) module and a 2-layer multi-layer perceptron (MLP) module. To ensure stable training, layer normalization (LN) layers are inserted before each MSA and MLP module, and a residual connection is applied after each module. The input for the Transformer block is formed by treating each voxel of the feature map along the channel dimension as a sequence. The W-MSA module adopts the same windowing operation as Swin Transformer~\citep{liu2021swin}, i.e., computing attention in the partitioned windows, instead of the whole feature maps. Specifically, in the W-MSA module, we split the feature map evenly into non-overlapping windows with a size of $4\times4\times4$ and compute self-attention within each local window. Meanwhile, to preserve the spatial information, we encode position information along three directions according to~\citep{carion2020end}. The final position encoding is obtained by concatenating the position encodings along these three directions. By adopting a hybrid architecture (consisting of alternating CNN and Transformer blocks), we can enhance the denoising network's capacity to capture long-range information while effectively keeping computational efficiency.

\subsection{Reconstruction Network}
By applying the Radon transform to the outputs of the denoising network, we can obtain the denoised images. Then, we use the reconstruction network (RecNet) $\mathbb{R}$ to reconstruct the SPET images $\hat{I}_{Std}$ from the denoised images. However, due to the nature of PET imaging, which reflects metabolic and biochemical activities rather than structural details, the inherent structural information in PET images is not as prominent as in other structural imaging modalities such as CT and MR. Thus, preserving and enhancing high-frequency structural information in the LPET to SPET reconstruction process poses a challenge. To address this, we incorporate the discrete wavelet transform (WT)~\citep{liu2018multi} into the reconstruction network. By dividing the PET image into several sub-band frequency spectrum maps, we can better reconstruct information of each sub-band, resulting in the reconstructed SPET image containing richer structural details. 
As shown in Figure~\ref{fig_framework} (b), we utilize discrete wavelet transform (WT) and inverse wavelet transform (IWT) in the reconstruction network $\mathbb{R}$ as a substitute for the traditional upsampling operations (such as zero-padding) and downsampling operations (such as max-pooling). Specifically, $\mathbb{R}$ includes an encoder and a decoder. First, the denoised image is passed through the encoder to generate 4096 sub-band frequency spectrum maps. Then, these frequency spectrum maps are utilized as the input of the decoder to reconstruct the SPET image $\hat{I}_{Std}$.

In the encoder, WT, instead of the pooling operation, is used to reduce the size of feature maps. Passing through a WT operation, an image/feature map is decomposed into eight sub-band feature maps of reduced size (e.g., a $256\times256\times160$ image will be decomposed into eight $128\times128\times80$ sub-band feature maps). We totally cascade four WT operations in the encoder, each followed by a fully CNN block consisting of four $3 \times 3\times 3$ convolutional layers (each with batch normalization (BN) and rectified linear unit (ReLU) operations). After passing through the encoder, the input denoised image will be decomposed into $4096$ feature maps. Considering that the relationship among different sub-bands may be useful to the reconstruction, we regard the sub-band feature maps as a concatenation along the channel dimension, rather than independent feature maps, when feeding into the CNN blocks.

In the decoder, correspondingly, four inverse wavelet transforms (IWTs) are cascaded for upsampling. Contrary to WT, IWT composes eight sub-band feature maps into one feature map/image of larger size (e.g., composing eight $128\times128\times80$ sub-band feature maps into one $256\times256\times160$ feature map). Each IWT is followed by a CNN block, similar to WT in the encoder. The only difference is that the last layer of the last CNN block has no BN and ReLU, which is adopted to predict the final result. That is,  through the decoder, the 4096 sub-band feature maps produced by the encoder will be finally composed into one image ($\hat{I}_{Std}$).

\subsection{Adversarial Network}
 In clinical practice, PET data is typically utilized in the image domain for diagnosis purposes, where information is obtained from images instead of sinograms or feature maps. As such, the main objective of our TriPLET is to produce high-quality SPET representations in the image domain, even if it involves multiple representations of PET data. To achieve this, we designed the adversarial networks (AdvNet) $\mathbb{A}$ in the image domain to evaluate the quality of the $\hat{I}_{Std}$ output from the reconstruction network.

As shown in Figure~\ref{fig_framework} (c), $\mathbb{A}$ takes two image pairs as input, including the fake pair and the real pair. The fake pair consists of the LPET image ${I}_{Low}$ and the predicted SPET image $\hat{I}_{Std}$, while the real pair contains the LPET image ${I}_{Low}$ and the actual SPET image ${I}_{Std}$. When being fed into the network, the images in each pair are concatenated along the channel dimension. The purpose of including LPET in the input of $\mathbb{A}$ is to impose an additional condition constraint on the discriminative process. In this way, we can reduce the difficulty of discrimination while ensuring the generator ($\mathbb{D}$ and $\mathbb{R}$) to produce a correct SPET image (corresponding to the structure of the inputted LPET, instead of an arbitrary SPET that just matches the real SPET distribution). The architecture of $\mathbb{A}$ is based on the traditional discriminator~\citep{isola2017image}. Specifically, it comprises four convolutional layers, with the first three using Leaky-ReLU activation functions and the last one using a sigmoid activation function to classify the input image pair as either real or fake.


\subsection{Loss Function}
While training our TriPLET, constraints in the projection, frequency, and images domains are all adopted, which are denoted as $\mathcal{L}_{P}$, $\mathcal{L}_{F}$, and $\mathcal{L}_{I}$, respectively. Using $\mathcal{L}_{P}$, $\mathcal{L}_{F}$, and $\mathcal{L}_{I}$ to supervise the training of TriPLET has two main benefits. On the one hand, our TriPLET consists of three sub-networks (i.e., denoising, reconstruction, and adversarial networks), and $\mathcal{L}_{P}$, $\mathcal{L}_{F}$, and $\mathcal{L}_{I}$ are designed specifically for these three sub-networks. Utilizing $\mathcal{L}_{P}$, $\mathcal{L}_{F}$, and $\mathcal{L}_{I}$ allows for better training of these sub-networks, thereby reducing the overall training difficulty of TriPLET. On the other hand, as mentioned in~\citep{abouhawwash2022evolutionary}, multiple learning objectives can enhance the overall quality of the reconstructed images from various perspectives, resulting in high-quality images that better meet the needs of practical applications. Specifically, $\mathcal{L}_{P}$ is performed in the projection domain, which is the mean square error (MSE) of the output of $\mathbb{D}$ against the SPET sinogram $S_{Std}$. The calculation of MSE is given by Equation~\ref{equation 5}. Hence, we can define $\mathcal{L}_{P}$ as below:
\begin{equation}
\begin{aligned}
 \mathcal{L}_{P} = \mathcal{L}_{MSE}(S_{Std}, \mathbb{D}(S_{Low})).
\end{aligned}
\label{equation 1}
\end{equation}

$\mathcal{L}_{F}$ is performed in the wavelet domain to impose constraints on all eight frequency spectrum maps that are output from the $\mathbb{R}$. Specifically, $\mathcal{L}_{F}$ utilize the focal frequency loss (FFL)~\citep{jiang2021focal}, which assigns varying weights to different elements of the frequency spectrum maps, i.e., larger weights for hard-to-reconstruct elements and smaller weights for easy-to-reconstruct elements. To achieve this, $\mathcal{L}_{F}$ introduces the weight matrices $W=[w_{1},w_{2},\cdots,w_{8}]$ with the same size as frequency spectrum maps that is dynamically determined by a nonuniform distribution on the current loss of each element during training. Let $F_{Std}=[f_{1},f_{2},\cdots,f_{8}]$ and $\hat{F}_{Std} = [\hat{f}_{1},\hat{f}_{2},\cdots,\hat{f}_{8}]$ be the frequency spectrum maps of actual SPET image and predicted SPET image, respectively, and then the value of the weight matrix $w_i$ at ($u,v,k$) can be calculated as:\\ 
\begin{equation}
\begin{aligned}
 w_i(u, v, k) = |f_{i}(u,v,k) -\hat{f}_{i}(u,v,k) |^{\alpha},
\end{aligned}
\label{equation 2}
\end{equation}
where $\alpha$ is the scaling factor for flexibility and set to $1$ in our implementation. Next, we normalize values of $w_i$ to fall within the range [$0$,$1$] to obtain $\overline{w}_i$, and perform a Hadamard product between the weight matrices $\overline{w}_i$ and the frequency distance matrices. This yields the complete form of $\mathcal{L}_{F}$:\\
\begin{equation}
\begin{aligned}
 \mathcal{L}_{F} = \sum_{i=1}^8( \overline{w}_i \odot \mathcal{L}_{MSE}(f_{i},\hat{f}_{i})).
\end{aligned}
\label{equation 3}
\end{equation}

$\mathcal{L}_{I}$ is performed in the image domain, considering two aspects. First, we calculate the MSE between the predicted SPET image $\hat{I}_{Std}$ and the actual SPET image $I_{Std}$, denoted as $\mathcal{L}_{I\_MSE}=\mathcal{L}_{MSE}(I_{Std},\hat{I}_{Std})$. Additionally, we incorporate the adversarial loss generated by the adversarial network $\mathbb{A}$, denoted as $\mathcal{L}_{I\_Adv}$. The formulation of $\mathcal{L}_{I\_Adv}$ is as follows:\\
\begin{equation}
\mathcal{L}_{I\_Adv}= (\mathbb{A}(I_{Low},I_{Std})-1)^2 + \mathbb{A}(I_{Low},\hat{I}_{Std})^2,
\label{equation 4}
\end{equation}
where $\mathbb{A}( , )$ represents the output probability of AdvNet. It serves as a measure of the likelihood or confidence assigned to a given input.




\subsection{Training Details}
We adopt a multi-stage training strategy to reduce the difficulty of training TriPLET, which divides the training process into three stages. In the first stage, $\mathbb{D}$ is independently supervised by $\mathcal{L}_{P}$. In the second stage, $\mathbb{D}$, $\mathbb{R}$, and $\mathbb{A}$ are co-supervised by the combination of $\mathcal{L}_{F}$ and $\mathcal{L}_{I}$ (including $\mathcal{L}_{I\_MSE}$ and $\mathcal{L}_{I\_Adv}$). It should be noted that in this stage, only the parameters of $\mathbb{R}$ and $\mathbb{A}$ are updated, while $\mathbb{D}$ is kept frozen. In the third stage,$\mathbb{D}$, $\mathbb{R}$, and $\mathbb{A}$ are co-supervised by the combination of $\mathcal{L}_{P}$, $\mathcal{L}_{F}$, and $\mathcal{L}_{I}$ and fine-tuned with a smaller learning rate. In the second and third stages, we utilize the GradNorm~\citep{chen2018gradnorm} technique to combine different loss functions, which allows for dynamic adjustment of the weights of the different losses based on their gradients during training.

The experiments are performed using the PyTorch platform and a single NVIDIA Tesla V100 GPU. Each of the three stages is trained for a total of 300 epochs using the Adam optimizer, and the learning rates for the three stages are set to $0.001$, $0.001$, $0.0001$, respectively.

\begin{figure}[!t]
\centering
\begin{overpic}[width=1\linewidth]{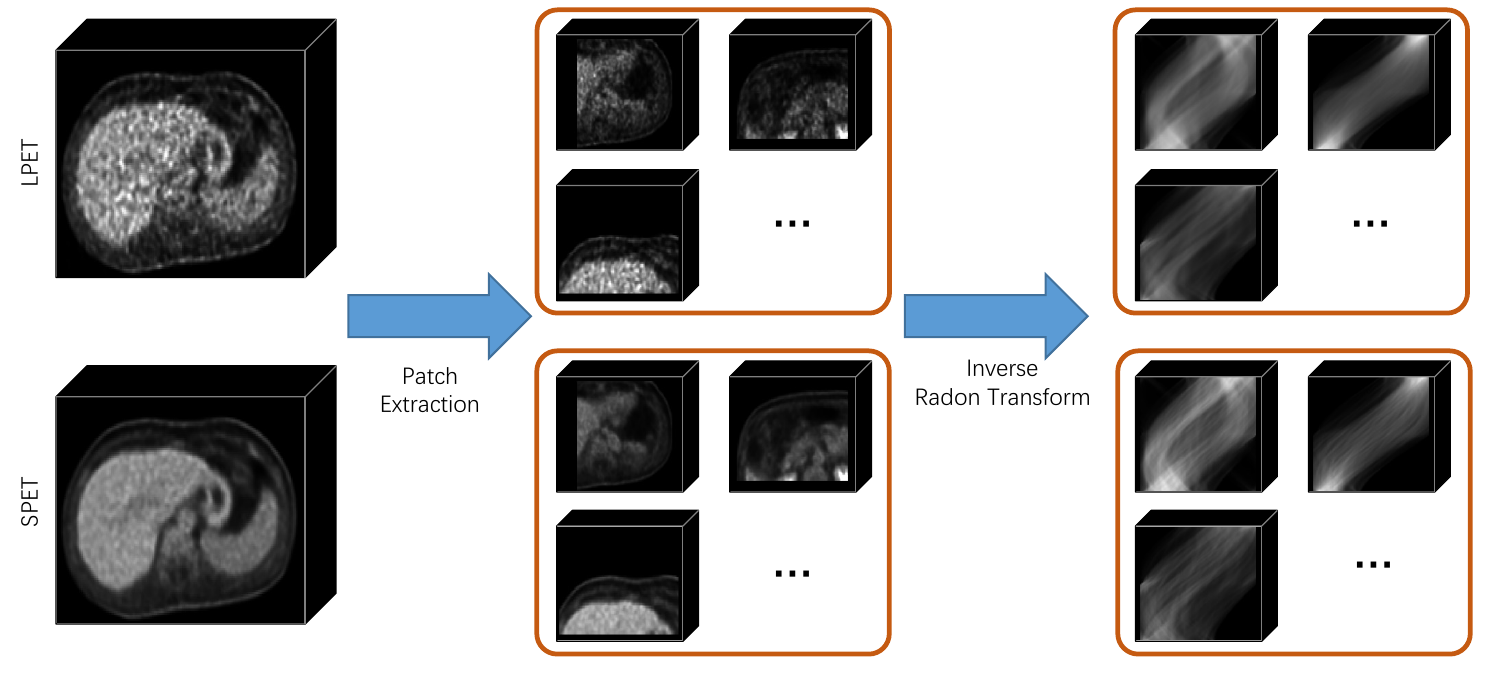}

    \end{overpic}

\centering 
\caption{Process of extracting image patches and sinogram patches from the whole PET image.}
\label{fig_patch_extraction}
\end{figure}

\begin{table}[!t]
\centering
\renewcommand\arraystretch{1.1}
\setlength\tabcolsep{8pt}

\caption{Physical characteristics of uEXPLORER scanner.}
\resizebox{70mm}{27mm}{
\begin{tabular}{l|c} 
\toprule[0.5mm]
Parameter     & Description   \\ 
\hline
Scintillator          & \makecell[c]{Lutetium-yttrium \\oxyorthosilicate} \\
Photodetectors          & Silicon photomultiplier \\
Crystal pitch and depth & $2.85 \times 2.85 \times 18.1$ mm \\ 
Total number of crystals & $ 564,480$  \\
AFOV  & $194$ cm \\
Detector ring diameter & $78.6$ cm \\
Transaxial ﬁeld of view & $68.6$ cm \\
TOF resolution &  $430$ ps\\
Energy resolution & $11.7$ \% \\
Number of lines of response & \makecell[c]{$\le 90$ billion before \\accounting for TOF}\\
Maximum axial (polar) angle & $57^{\circ}$\\
\bottomrule[0.4mm]
\end{tabular}}
\label{table_characteristics}
\end{table}

\section{Experiments}
\subsection{Data Description}
For the evaluation of TriPLET, we collect a dataset of 70 paired chest-abdomen SPET and LPET images from the uEXPLORER PET/CT scanner~\citep{zhang2020total} whose relevant parameters are given in Table~\ref{table_characteristics}. Paired LPET and SPET images are acquired using the list mode of the scanner with an injection of $256$ MBq of [$^{18}$F]-FDG injection. The SPET images are reconstructed using 1200s data between 60 minutes and 80 minutes after injection by ordered-subsets expectation maximization (OSEM) algorithm~\citep{mehranian2020model}, while the corresponding 1/10th LPET images are simultaneously reconstructed using 120s data uniformly sampled from the 1200s data.

In the preprocessing stage, we uniformly resample all images to a voxel spacing of $1\times1\times1~\text {mm}^{3}$ and a resolution of $256 \times 256 \times 160$. We filter each PET image's values (i.e., only retaining values within the 5\% to 95\% range) to eliminate outliers, thereby removing their influence on subsequent tasks. Then, the intensity range of the images is normalized to [$0,1$] using Z-score normalization. To increase the number of training samples and reduce the dependence on GPU memory, we extract overlapping patches of size $96 \times 96 \times 96$ from each whole PET image. As shown in Figure~\ref{fig_patch_extraction}, all patch extraction operations are performed in the image domain, and the corresponding sinogram patches are converted from the extracted image patches through inverse Radon transform. The sinogram and image can be losslessly converted between each other through Radon transform and inverse Radon transform~\citep{toft1996radon}. Hence, this method of obtaining sinogram patches does not introduce additional errors and ensures consistency between image patches and sinogram patches. It should be noted that our patch extraction operation is directly applied to PET images in the image domain and cannot be directly applied to PET sinograms. However, in theory, when the input is a sinogram, we can also extract patches from the sinogram and then use the Radon transform to obtain the corresponding image patches. Additionally, we did not collect PET raw data (i.e., sinograms). The sinograms indirectly obtained from reconstructed PET images may indeed differ from the real PET raw data. Nevertheless, our method can still be applied to real PET raw data, although the performance might exhibit some variation. From each subject, 40 patches covering all regions-of-interest are extracted, resulting in a total of 2800 training samples.

To ensure stable results and minimize randomness, we adopted a five-fold cross-validation. In each fold, samples from 56 subjects are used for training, while the remaining subjects are used for testing. Mean and standard deviation are calculated across the five folds to report the overall performance.

\subsection{Evaluation Metrics}
In our evaluation of TriPLET, we employ three commonly used metrics in reconstruction tasks to quantitatively assess the results, including structural similarity index (SSIM)~\citep{wang2004image}, relative root mean squared error (rRMSE), and peak signal-to-noise ratio (PSNR).

\begin{table*}
\centering
\renewcommand\arraystretch{1.5}
\setlength\tabcolsep{7pt}
\caption{Performance of SPET images reconstructed using five different combinations of loss functions and network components, measured by PSNR, SSIM, and rRMSE. The symbol $^{\ast}$ indicates  the significant difference  between the performance of our approach and the optimal performance among other methods, at $p$-value \textless 0.05 level.}
\begin{tabular}{c|l|ccc|c|c|c}
\toprule[0.5mm]
 Method &  Network      & $\mathcal{L}_{P}$ & $\mathcal{L}_{F}$ & $\mathcal{L}_{I}$ & \multicolumn{1}{c|}{PSNR [dB]$\uparrow$ } & \multicolumn{1}{c|}{SSIM $\uparrow$} & \multicolumn{1}{c}{rRMSE $\downarrow$}  \\ 
\hline
\uppercase\expandafter{\romannumeral1}&  U-Net          &   \rule[4pt]{0.2cm}{0.09em}  &  \XSolidBrush  &    \Checkmark        &$23.367~\pm~1.372$ & $0.983~\pm~ 0.012$ & $0.792~\pm~ 0.557$\\
\hline
\uppercase\expandafter{\romannumeral2} &  RecNet         &  \rule[4pt]{0.2cm}{0.09em}  &  \XSolidBrush &  \Checkmark &   $23.865 ~\pm~ 1.486$ & $0.985~\pm~ 0.009$  & $0.468~\pm~ 0.365$\\
\uppercase\expandafter{\romannumeral3} &  RecNet     & \rule[4pt]{0.2cm}{0.09em}   &  \Checkmark &  \Checkmark &      $24.968 ~\pm~ 0.973$ & $0.988~\pm~ 0.007$  & $0.413~\pm~ 0.327$\\
\hline
\uppercase\expandafter{\romannumeral4} &DenNet + RecNet   & \Checkmark  &  \XSolidBrush & \Checkmark  &  $25.341 ~\pm~ 0.845$ & $0.990~\pm~ 0.006$ & $0.316~\pm~ 0.243$\\
\uppercase\expandafter{\romannumeral5} (ours) & DenNet + RecNet &  \Checkmark & \Checkmark  & \Checkmark  &$\bm{25.932}~\pm~\bm{0.671}^{\ast}$ & $\bm{0.992}~\pm~\bm{0.004}$  & $\bm{0.296}~\pm~\bm{0.213}^{\ast}$  \\ [3pt]   \bottomrule[0.4mm]              
\end{tabular}
\vspace{2mm}
\label{table1}
\end{table*}

 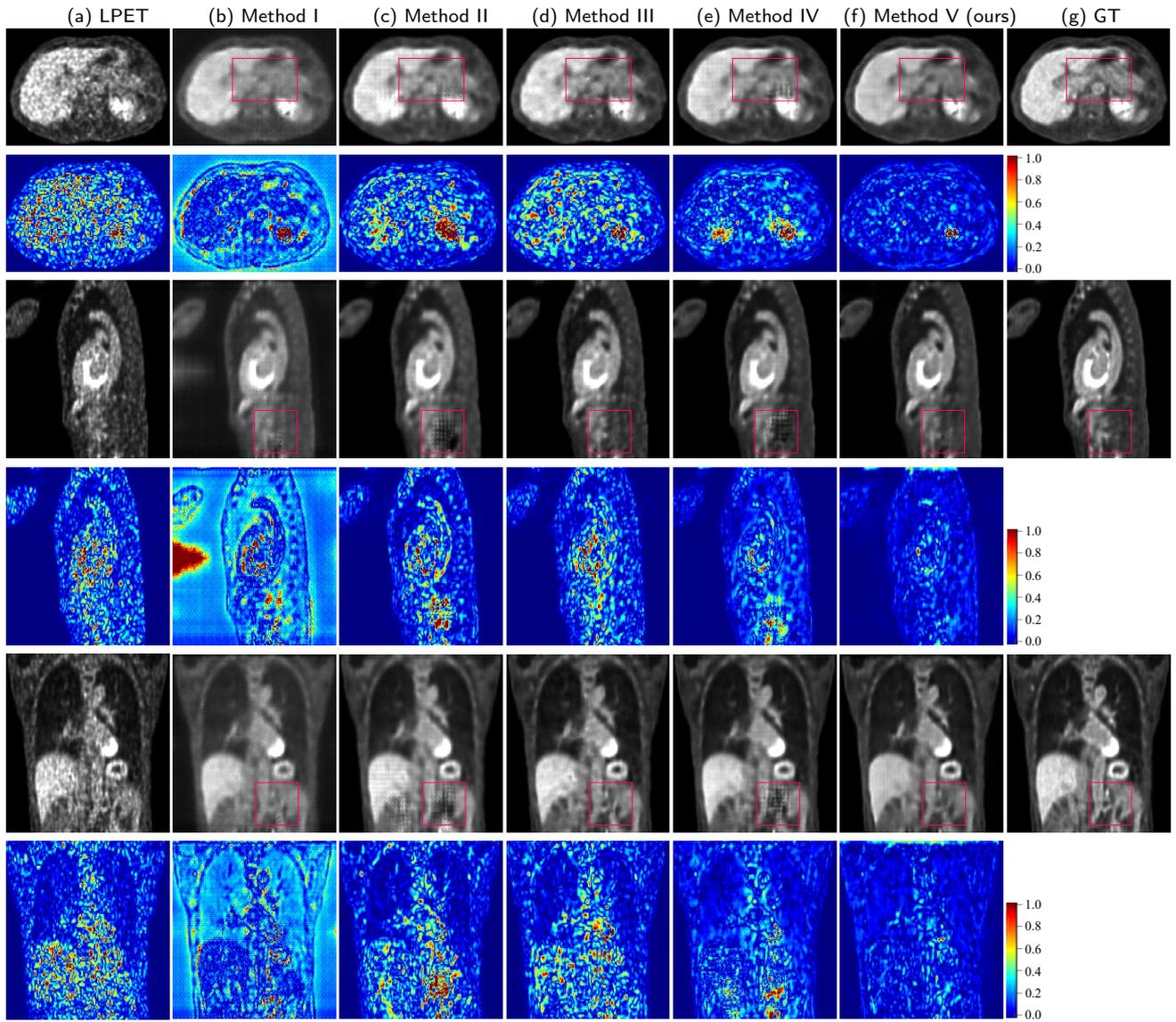
\begin{figure*}[!t]
 \setlength{\abovecaptionskip}{0.1cm}
\setlength{\belowcaptionskip}{-0.4cm}
\centering
\begin{overpic}[width=1\linewidth]{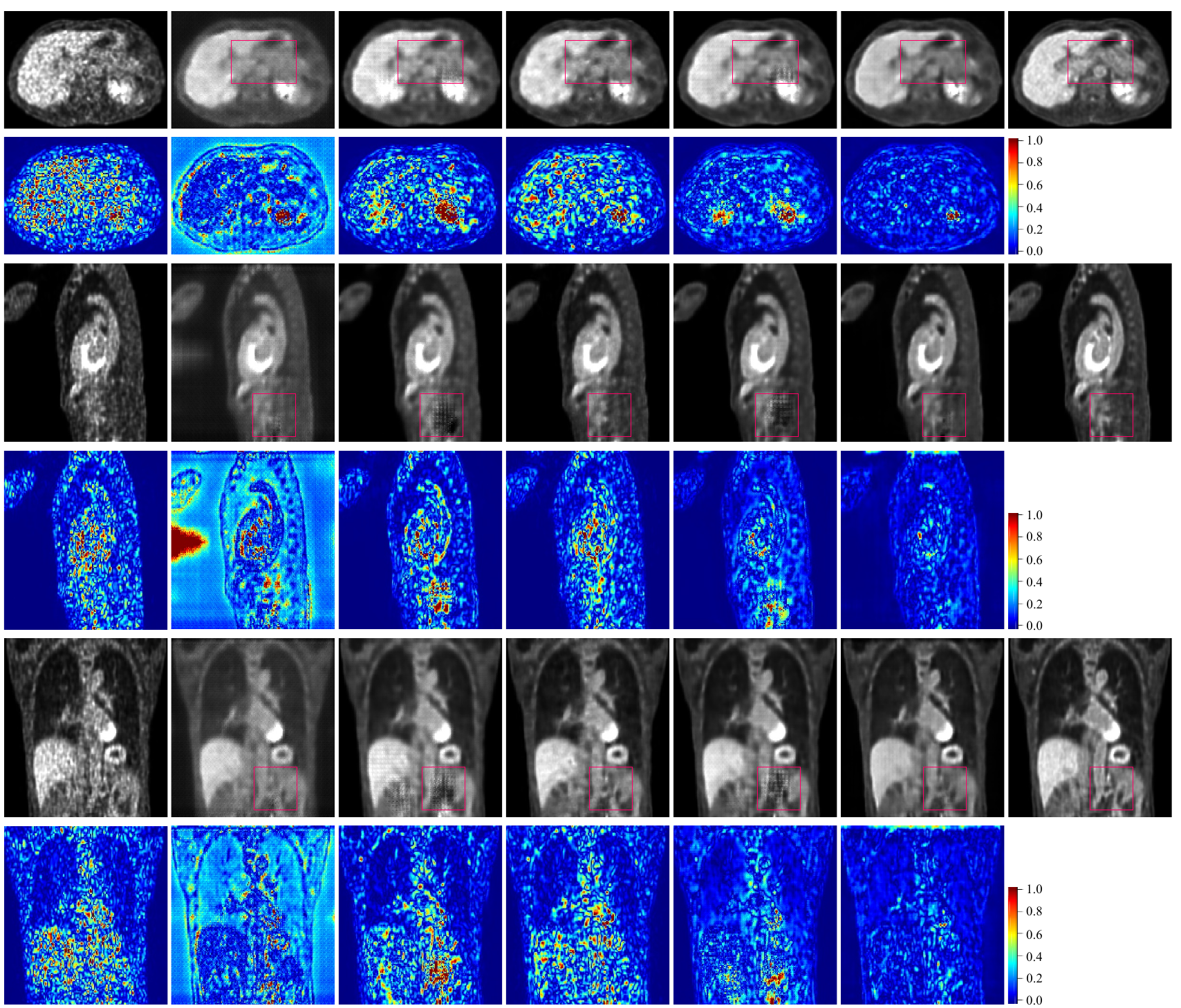}

         \put(5.5, 85){\footnotesize (a) LPET}
        \put(17.5, 85){\footnotesize (b) Method \uppercase\expandafter{\romannumeral1}}
        \put(31.5, 85){\footnotesize (c) Method \uppercase\expandafter{\romannumeral2}}
        \put(45, 85){\footnotesize (d) Method \uppercase\expandafter{\romannumeral3}}
        \put(59, 85){\footnotesize (e) Method \uppercase\expandafter{\romannumeral4}}
        \put(71.5, 85){\footnotesize (f) Method \uppercase\expandafter{\romannumeral5} (ours)}
        \put(90, 85){\footnotesize (g) GT}

    \end{overpic}
\centering
\caption{Visual comparison of SPET images produced using five different combinations of loss functions and network components shown in Table 1. The LPET image is shown on the left, followed by the results produced by the five different methods indicated in Table 1 (2nd-5th columns), and the ground truth (GT) in the last column. The corresponding difference maps between the produced results and the ground truth (GT) are depicted in the 2nd (axial view), 4th (sagittal view), and 6th (coronal view) rows. Red boxes show areas for detailed comparison.}
\label{fig_ablation}
\end{figure*}

Specifically, the SSIM metric is particularly useful for assessing visual image quality, as it takes image structure into account. A high SSIM value indicates good image quality, with a range of $0\leq \text{SSIM} \leq1$. An SSIM value of 1 signifies that the compared image is identical to the ground-truth image.

The rRMSE metric evaluates the level of agreement between two sets of measurements by assessing the root mean squared error relative to a reference. It quantifies the accuracy of the reconstruction, with lower rRMSE values indicating higher agreement with the ground truth,

\begin{equation}
\begin{aligned}
&\text{MSE} = \frac{1}{N}\sum_{i=1}^N(x_i-y_i)^2,\\
&\text{rRMSE} = \frac{\sqrt{\text{MSE}}}{\overline{y}},
\end{aligned}
\label{equation 5}
\end{equation}
where $x$ is the predicted image, $y$ is the ground-truth image, $N$ is the number
of image voxels, and $\overline{y}$ is the all-voxel-averaged value of the ground-truth image. 

And PSNR metric measures the ratio of the maximum possible power of a signal to the power of corrupting noise, with higher PSNR values indicating better agreement between the predicted image and the ground-truth image,
\begin{equation}
\text{PSNR} = 20\cdot\log_{10}(\frac{y_{max}}{\sqrt{\text{MSE}}}).
\label{equation 6}
\end{equation}
$y_{max}$ represents the maximum value of the ground-truth image.

We use difference maps to display the qualitative results. Difference maps are obtained by subtracting the reconstructed SPET images (i.e., the actual SPET) from the GT and then taking the absolute value. In the difference maps, larger values (i.e., darker colors) indicate a larger difference between the reconstructed SPET image and the GT, while smaller values (i.e., lighter colors) indicate a smaller difference between the reconstructed SPET image and the GT.

\subsection{Ablation Study of Key Components}
To verify the contributions of different network components and loss functions of our proposed TriPLET, we perform an ablation experiment to compare the following five methods: \uppercase\expandafter{\romannumeral1}) U-Net supervised by $\mathcal{L}_{I}$; \uppercase\expandafter{\romannumeral2}) $\mathbb{R}$ supervised by $\mathcal{L}_{I}$; \uppercase\expandafter{\romannumeral3}) RecNet supervised by $\mathcal{L}_{F}$ and $\mathcal{L}_{I}$; \uppercase\expandafter{\romannumeral4}) DenNet and RecNet supervised by $\mathcal{L}_{P}$ and $\mathcal{L}_{I}$; \uppercase\expandafter{\romannumeral5}) DenNet and RecNet supervised by $\mathcal{L}_{P}$, $\mathcal{L}_{F}$, and $\mathcal{L}_{I}$.
It is noteworthy that the production of $\mathcal{L}_{I}$ is dependent on the adversarial networks $\mathbb{A}$, therefore all the methods discussed above involve $\mathbb{A}$. Furthermore, Methods \uppercase\expandafter{\romannumeral1}, \uppercase\expandafter{\romannumeral2}, and \uppercase\expandafter{\romannumeral3} aim to reconstruct SPET images from LPET images, whereas Methods \uppercase\expandafter{\romannumeral4} and \uppercase\expandafter{\romannumeral5} reconstruct SPET images from LPET sinograms. All methods use the same experimental settings, and their quantitative and qualitative results are provided in Table~\ref{table1} and Figure~\ref{fig_ablation}, respectively.

\begin{table*}[!t]
\centering
\renewcommand\arraystretch{1.3}
\setlength\tabcolsep{8pt}
\caption{Quantitative comparison between our TriPLET and state-of-the-art SPET reconstruction methods, measured by PSNR, SSIM, and rRMSE. The symbol $^{\ast}$ indicates the significant difference between the performance of our approach and the optimal performance among other methods, at $p$-value \textless 0.05 level.}
\begin{tabular}{l|c|c|c} 
\toprule[0.5mm]
Method         & \multicolumn{1}{c|}{PSNR [dB]$\uparrow$ } & \multicolumn{1}{c|}{SSIM $\uparrow$} & \multicolumn{1}{c}{rRMSE $\downarrow$}  \\ 
\hline
cGAN        & $23.383~\pm~1.364$ & $0.982~\pm~ 0.013$ &  $0.604~\pm~ 0.483$\\
LA-GAN       &$23.351~\pm~1.725$  & $0.984~\pm~ 0.012$ & $0.767~\pm~ 0.551$  \\
Sinogram-Net    &$23.528~\pm~ 1.634$  & $0.986~\pm~ 0.009$ &  $0.541~\pm~ 0.416$  \\
Trans-GAN    &$23.852~\pm~ 1.522$  & $0.985~\pm~ 0.011$ &  $0.452~\pm~ 0.372$  \\
AR-GAN    &$25.217~\pm~ 0.853$  & $0.988~\pm~ 0.007$ &  $0.377~\pm~ 0.359$  \\
TriDoRNet &$25.892~\pm~ 0.784$  & $0.991~\pm~ 0.006$ &  $0.325~\pm~ 0.286$  \\
TriPLET (ours) &$\bm{25.932}~\pm~\bm{0.671}^{\ast}$ & $\bm{0.992}~\pm~\bm{0.004}^{\ast}$  & $\bm{0.296}~\pm~\bm{0.213}^{\ast}$  \\
\bottomrule[0.4mm]
\end{tabular}
\vspace{2mm}
\label{table2}
\end{table*}

 \begin{figure*}[!t]
 \setlength{\abovecaptionskip}{0.1cm}
\setlength{\belowcaptionskip}{-0.4cm}
\centering
\begin{overpic}[width=1\linewidth]{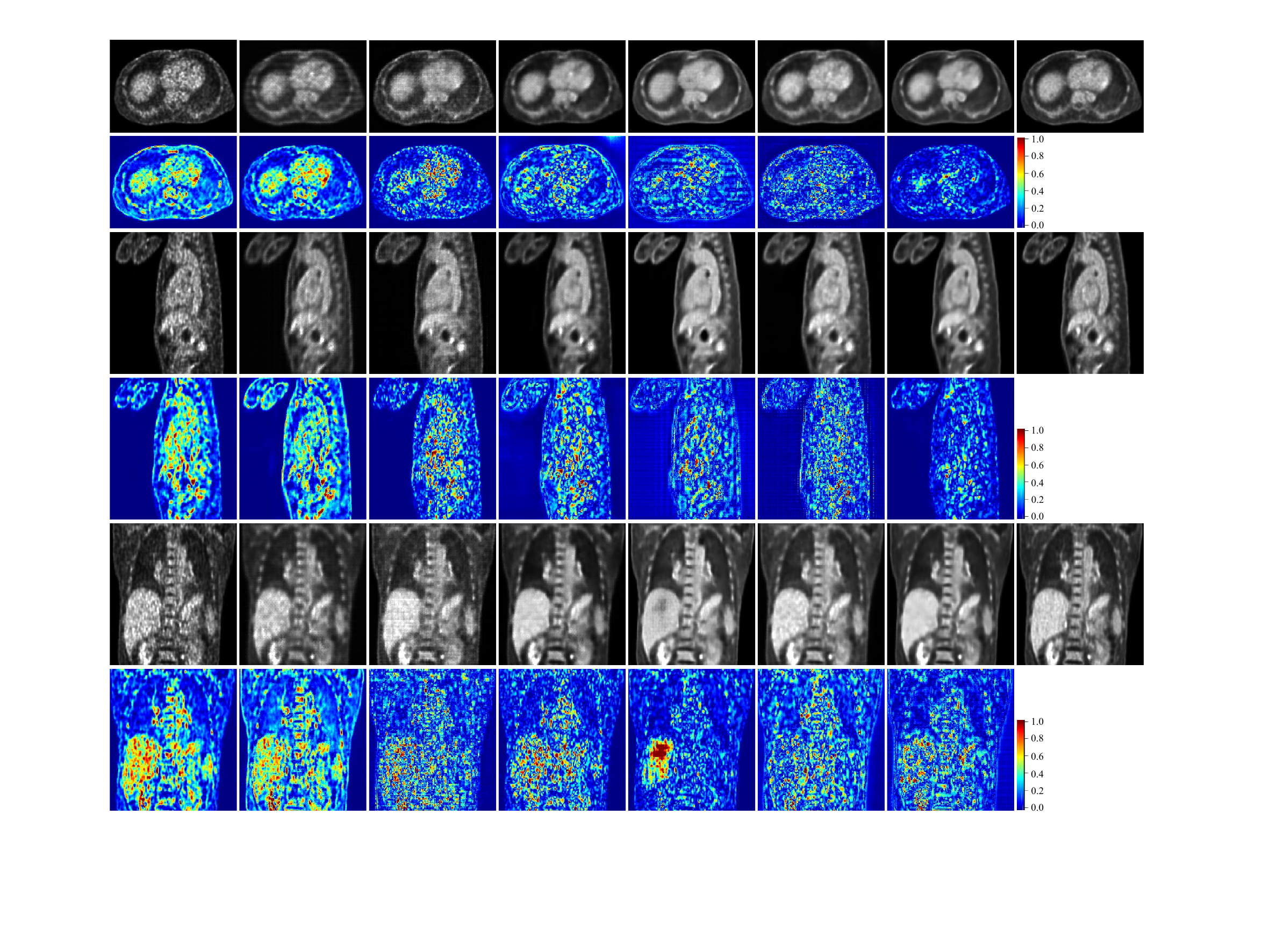}
         \put(4, 75){\footnotesize (a) LPET}
        \put(16, 75){\footnotesize (b) cGAN}
        \put(27.5, 75){\footnotesize (c) LA-GAN}
        \put(38, 75){\footnotesize (d) Sinogram-Net}
        \put(51.7, 75){\footnotesize (e) Trans-GAN}
        \put(64.4, 75){\footnotesize (f) AR-GAN}
        \put(75.5, 75){\footnotesize (g) TriPLET (ours)}
        \put(90.5, 75){\footnotesize (h) GT}
    \end{overpic}
    \vspace{-1mm}
\centering
\caption{Visual comparison of SPET images generated by six different methods. The arrangement from left to right includes the LPET images, results obtained from five alternative comparison methods (2nd to 6th columns), our TriPLET (7th column), and the GT (SPET image). Additionally, the 2nd (axial view), 4th (sagittal view), and 6th (coronal view) rows display the corresponding difference maps between the generated results and the GT.}
\label{fig_comparision}
\end{figure*}
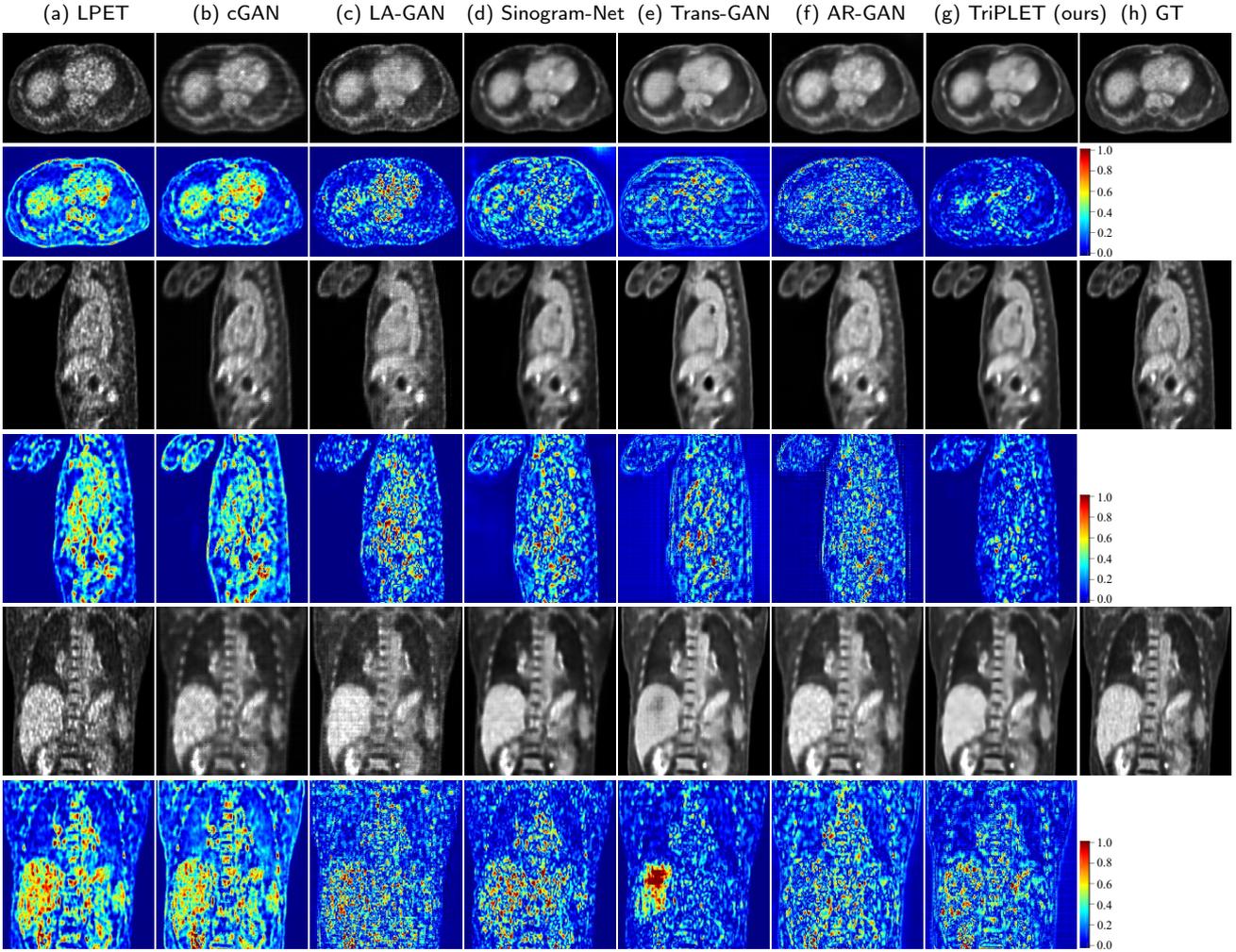

 \textbf{Benefits of Wavelet Transform:} The convolutional and pooling layers in convolutional neural networks (CNNs) tend to smooth and downsample voxels when processing images~\citep{huang2022deep}. This can result in detailed information loss and also the over-smoothed predicted images, which is particularly problematic for PET images that inherently lack structural details. To overcome this challenge, we incorporate discrete wavelet transform into the reconstruction network $\mathbb{R}$ to decompose the PET images into multiple sub-band frequency spectrum maps. Our designed $\mathbb{R}$ can decompose the high-frequency components in PET images separately through WT during reconstruction, and it is equipped with a specially designed loss function ($\mathcal{L}_{F}$) to supervise the reconstruction of the high-frequency components. In this way, we can enable the network to focus more on the high-frequency components during the reconstruction process, thus achieving better preservation and reconstruction of the high-frequency components.


Based on experimental results, it can be observed that $\mathbb{R}$ integrated with discrete wavelet transform performs better in SPET reconstruction than U-Net. Specifically, as depicted in Figure~\ref{fig_ablation}, the SPET images predicted by Methods \uppercase\expandafter{\romannumeral2} and \uppercase\expandafter{\romannumeral3} exhibit more distinct boundaries, compared to Method \uppercase\expandafter{\romannumeral1}, particularly in the unclear tissue regions, as highlighted by red bounding boxes in the first row. Additionally, the improvement is evidenced by superior quantitative metrics from Methods \uppercase\expandafter{\romannumeral2} and \uppercase\expandafter{\romannumeral3} that outperform the metrics from Method \uppercase\expandafter{\romannumeral1}, as shown in Table~\ref{table1}. Specifically, Methods \uppercase\expandafter{\romannumeral2} and \uppercase\expandafter{\romannumeral3} exhibit an improvement of $0.498$/$1.601$ dB, $0.002$/$0.005$, and $0.324$/$0.379$ in PSNR, SSIM, and rRMSE, respectively. This serves as further evidence of the effectiveness for the discrete wavelet transform in improving SPET reconstruction.

\textbf{Denoising for Sinogram:} In TriPLET, the mapping from LPET to SPET is formulated as both denoising and intensity distribution predictions. By cascading a specialized denoising network $\mathbb{D}$ to denoise the LPET sinogram before feeding it into the reconstruction network $\mathbb{R}$, we can reduce the burden of the reconstruction network and effectively prevent the propagation and amplification of noise in the subsequent processing, thus achieving better performance in SPET reconstruction.

The effectiveness of denoising for sinograms is demonstrated in the experimental results as presented in Table~\ref{table1} and Figure~\ref{fig_ablation}. Specifically, as shown in Figure~\ref{fig_ablation}, both $\mathbb{D}+\mathbb{R}$ methods (Methods \uppercase\expandafter{\romannumeral4} and \uppercase\expandafter{\romannumeral5}), which perform denoising first and then reconstruction, achieve better results than the only $\mathbb{R}$ methods (i.e., Methods \uppercase\expandafter{\romannumeral2} and \uppercase\expandafter{\romannumeral3}. Notably, the SPET images ((e) and (f))reconstructed using $\mathbb{D}+\mathbb{R}$ methods have less noise than those ((c) and (d)) by the only $\mathbb{R}$ methods. In addition, Table~\ref{table1} shows that $\mathbb{D}+\mathbb{R}$ methods achieve better performance than the only $\mathbb{R}$ methods in terms of not only PSNR, but also SSIM, and rRMSE. This verifies that an initial denoising (in the projection domain) step with advanced methods is helpful to boost the reconstruction results, and the cascaded framework of our TriPLET is effective.

\textbf{Usefulness of Frequency Loss:}
In the reconstruction network $\mathbb{R}$, the discrete wavelet transform decomposes the PET image into a series of frequency spectrum maps with different sub-bands. To ensure proper reconstruction of information in each sub-band, we design a specific constraint $\mathcal{L}_{F}$ in the wavelet domain for the frequency spectrum maps outputted from $\mathbb{R}$.

Comparing the Methods \uppercase\expandafter{\romannumeral3} and \uppercase\expandafter{\romannumeral2}, or Methods \uppercase\expandafter{\romannumeral5} and \uppercase\expandafter{\romannumeral4}, $\mathbb{R}$ or $\mathbb{D}+\mathbb{R}$ can achieve better reconstruction results with the use of additional supervision $\mathcal{L}_{F}$ designed in the wavelet domain. Specifically, based on visual results in the second and third rows of Figure~\ref{fig_ablation}, it can be observed that the results ((c) and (e)) obtained without $\mathcal{L}_{F}$ exhibit artifacts in certain regions, as highlighted by red boxes. In contrast, the results ((d) and (f)) obtained with the inclusion of $\mathcal{L}_{F}$ are artifact-free, indicating that the direct supervision of the frequency spectrum map in the wavelet domain effectively prevents errors from propagating to the image domain. As a result, the final prediction image, which incorporates information from all sub-bands, is visually better with overall improved quality. Additionally, quantitative results in Table~\ref{table1} show that the former methods achieve boosted values of PSNR, SSIM, and rRMSE. These findings suggest that adopting the constraint $\mathcal{L}_{F}$ from the wavelet domain can enhance the performance of our WT-based reconstructed network.

 Finally, $\mathbb{D}+\mathbb{R}$ with supervision in all three domains (Method \uppercase\expandafter{\romannumeral5} achieves the best results (i.e., with the lightest color difference maps in Figure~\ref{fig_ablation} and the best PSNR, SSIM, and rRMSE values in Table~\ref{table1}). This proves that all of our proposed loss functions and network components contribute to the final performance.

\subsection{Comparison Experiment}
We further conducted a comprehensive comparison between our proposed TriPLET and several state-of-the-art SPET enhancement methods, including cGAN~\citep{20183D}, LA-GAN~\citep{20193D}, Sinogram-Net~\citep{feng2020rethinking}, Transformer-GAN~\citep{luo20213d}, AR-GAN~\citep{luo2022adaptive}, and TriDoRNet~\citep{jiang2023tridornet}. Specifically, cGAN, LA-GAN, Transformer-GAN, and AR-GAN are the GAN-based methods that estimate SPET images from LPET images, while Sinogram-Net and TriDoRNet are a coupled network that directly reconstructs SPET images from LPET sinograms. The quantitative and qualitative results are provided in Table~\ref{table2} and Figure~\ref{fig_comparision}, respectively.

\textbf{Quantitative Comparison:} 
Table~\ref{table2} presents the quantitative results obtained by different methods, measured in terms of PSNR, SSIM, and rRMSE. Our proposed approach demonstrates superior performance in all aspects. Specifically, when compared to cGAN, which yields the poorest results, our approach achieves significant improvements. We observe an increase of 2.549 dB in PSNR, a 0.010 increase in SSIM, and a reduction of 1.689 in rRMSE. 
Additionally, our method achieves better performance than TriDoRNet, proving that our improvements based on TriDoRNet are effective. Furthermore, statistical analysis using paired t-tests confirms that the performance of our TriPLET method is significantly better than the optimal performance achieved by other methods. This significance is indicated by the obtained $p$-value (\textless 0.05) for all evaluation metrics.

These findings provide strong evidence of the effectiveness and superiority of our proposed TriPLET approach in enhancing SPET images, as it outperforms other state-of-the-art methods in terms of both objective measures (PSNR, SSIM, rRMSE) and statistical significance.


\textbf{Qualitative Comparison:}
The visual results of the comparison methods are presented in Figure~\ref{fig_comparision}. It is evident that our approach excels in generating SPET images with several notable characteristics. Firstly, our method achieves the best performance as it can reconstruct SPET images that are closest to the GT, evidenced by having the lightest colored difference maps. Additionally, our method produces SPET images with the clearest boundaries and exhibits superior performance in areas that are difficult to reconstruct (e.g., the bronchial areas shown in the 5th row of Figure~\ref{fig_comparision}). In the difference maps, the corresponding boundaries and bronchial areas of our method have the lightest colors. Moreover, our approach effectively minimizes noise, resulting in cleaner images that enhance the visibility of important biological details. 

Collectively, these visual comparisons provide substantial evidence supporting the superiority of our approach over state-of-the-art methods. Our method consistently produces SPET images with clear boundaries, reduced noise, enhanced details, and minimal differences from the ground truth, further affirming its superior performance in SPET image generation.


\section{DISCUSSION}
\subsection{External Validation}
Further, we explore the performance of our approach across different PET datasets through external validation. Specifically, we source data from the Ultra Low Dose PET Imaging Challenge Dataset\footnote{\url{https://ultra-low-dose-pet.grand-challenge.org/Dataset/}}, which is a public dataset that includes paired SPET and corresponding LPET data acquired from Siemens Biograph Vision Quadra and United Imaging uEXPLORER scanners. From this dataset, we collect 40 pairs of SPET and their corresponding 1/10th LPET data as the external validation set. During external validation, all methods are trained on the internal dataset and then directly tested on the external validation set, with the corresponding SPET reconstruction results presented in Table~\ref{table3}. The results indicate that, even on the external validation dataset, our TriPLET outperforms other SPET reconstruction methods, further validating the effectiveness of our TriPLET. Moreover, our TriPLET shows the smallest performance difference between the external validation dataset and the internal dataset, indicating that compared to other methods, our TriPLET possesses the best robustness across various PET datasets.

\begin{table*}[!t]
\centering
\renewcommand\arraystretch{1.3}
\setlength\tabcolsep{8pt}

\caption{Quantitative comparison of external validation, measured by PSNR, SSIM, and rRMSE. The symbol $^{\ast}$ indicates the significant difference between the performance of our approach and the optimal performance among other methods, at $p$-value \textless 0.05 level.}
\begin{tabular}{l|c|c|c} 
\toprule[0.5mm]
Method         & \multicolumn{1}{c|}{PSNR [dB]$\uparrow$ } & \multicolumn{1}{c|}{SSIM $\uparrow$} & \multicolumn{1}{c}{rRMSE $\downarrow$}  \\ 
\hline
cGAN        & $22.882~\pm~1.753$ & $0.980~\pm~ 0.014$ &  $0.782~\pm~ 0.653$\\
LA-GAN       &$23.123~\pm~1.832$  & $0.982~\pm~ 0.013$ & $0.723~\pm~ 0.554$  \\
Sinogram-Net    &$23.132~\pm~ 1.852$  & $0.983~\pm~ 0.011$ &  $0.663~\pm~ 0.521$  \\
Trans-GAN    &$23.631~\pm~ 1.635$  & $0.984~\pm~ 0.009$ &  $0.496~\pm~ 0.475$  \\
AR-GAN    &$24.524~\pm~ 0.972$  & $0.985~\pm~ 0.008$ &  $0.397~\pm~ 0.462$  \\
TriDoRNet &$24.643~\pm~ 0.913$  & $0.987~\pm~ 0.006$ &  $0.348~\pm~ 0.315$  \\
TriPLET (ours) &$\bm{25.146}~\pm~\bm{0.862}^{\ast}$ & $\bm{0.989}~\pm~\bm{0.005}^{\ast}$  & $\bm{0.312}~\pm~\bm{0.256}^{\ast}$  \\
\bottomrule[0.4mm]
\end{tabular}
\vspace{2mm}
\label{table3}
\end{table*}

\subsection{Curse of Dimensionality}
The curse of dimensionality~\citep{e2022training} is a common challenge in data analysis, machine learning, and statistics. It describes the phenomenon where the amount of data samples required for analyzing data or training models grows exponentially with the increase in dimensions. The curse of dimensionality also exists in three-dimensional medical image analysis tasks, including ours. Three-dimensional medical images typically consist of a vast number of voxels, with each voxel representing a feature dimension. This results in extremely high data dimensions, and each dimension may hold important medical information, complicating tasks such as image generation and reconstruction.

In our work, we adopt several strategies to combat the curse of dimensionality. These include: 1) Regularization. We utilize Dropout techniques during the training process to randomly omit (i.e., set to zero) a portion of the neurons in the network. This approach can be considered a model averaging technique that helps mitigate the risk of overfitting and enhances the model's ability to generalize; 2) Data dimensionality reduction. Our framework incorporates an encoder-decoder structure (i.e., sub-network $\mathbb{R}$). Through the encoder, we can effectively reduce the dimensionality of the input data while preserving essential information; 3) The multi-stage training strategy, where we divide the training of TriPLET into three stages. In this way, we can significantly reduce the number of network parameters that need to be trained in each stage, thus reducing the reliance on training data and computational resources; 4) Data augmentation and five-fold cross-validation. During the data preprocessing stage, we extract 40 overlapping patches from each whole PET image, increasing the number of data samples from 70 to 2800. This not only expands our dataset but also reduces the network input size, thereby lessening the dependence on GPU memory. Moreover, the use of five-fold cross-validation during training allows us to make more efficient use of the augmented data.

\subsection{The Impact of Data Diversity and Quality}
The quality and diversity of data significantly impact tasks involving the reconstruction and generation of medical images. Factors such as noise, outliers, and a lack of data diversity can lead to poor model training outcomes, which in turn can affect the accuracy and reliability of the generated or reconstructed images.

In our work, we have employed a series of strategies to mitigate the impacts of these factors. Firstly, during the data preprocessing phase, we filter each PET image's values (i.e., only retaining values within the 5\% to 95\% range) to eliminate outliers, thereby removing their influence on subsequent tasks. We then normalize the range of all image values to [0,1] through z-score normalization to further eliminate the interference of noise and outliers. Secondly, by extracting 40 overlapping patches from each whole PET image, we increase the data samples from 70 to 2800. This method significantly enhances data diversity, enabling the trained model to possess stronger generalization. Lastly, in the model design phase, based on the principles of PET imaging, we decompose the task of mapping from LPET to SPET into two simpler tasks (namely, denoising and reconstruction tasks). In this way, we can more stably and robustly predict SPET from LPET, and the preliminary denoising network can further reduce the impact of noise on our task.
Through the above strategies, we effectively minimize the impact of data quality and diversity on our task. Using the same data to train models, our proposed TriPLET achieves superior SPET reconstruction results compared to other methods, with the results provided in Table~\ref{table2} and Figure~\ref{fig_comparision}.

\section{CONCLUSION}
In this work, we develop a novel end-to-end TriPle-domain LPET EnhancemenT (TriPLET) framework for reconstructing SPET Images from LPET sinograms, by considering a hybrid denoising-and-reconstruction process and simultaneously taking the advantages of projection, wavelet, and image domains into the enhancement procedure. Specifically, TriPLET couples three networks to sequentially perform sinogram denoising in the projection domain, SPET reconstruction in the wavelet domain, and image evaluation in the image domain. Furthermore, we specifically design three loss functions respective to the three domains for supervising the training of TriPLET. Through experiments conducted on real human chest-abdomen PET data, we demonstrate that TriPLET is designed effectively and can also achieve the best quantitative and qualitative results than state-of-the-art methods.




\section*{Declaration of Competing Interest}
The authors declare that they have no known competing financial interests or personal relationships that could have appeared to influence the work reported in this paper.

\section*{CRediT Authorship Contribution Statement}
\textbf{Caiwen Jiang:} Conceptualization, Methodology, Software, Writing - original draft, Validation. \textbf{Mianxin Liu:} Investigation, Conceptualization, Writing - review \& editing. 
\textbf{Kaicong Sun:} Investigation, Supervision, Writing - review \& editing. 
\textbf{Dinggang Shen:} Conceptualization, Supervision, Project administration, Writing - review \& editing, Funding acquisition.

\section*{Ethical and Data Availability Statement}
This work involved human subjects or animals in its research. Approval
of all ethical and experimental procedures and protocols was granted by the
Institute Research Medical Ethics Committee of Zhongshan Hospital, Fudan
University.

\section*{Acknowledgments}
This work was supported in part by the National Natural Science Foundation of China under Grant 62131015 and 82394432, the Science and Technology Commission of Shanghai Municipality (STCSM) under Grant 21010502600, and the Key Research and Development Program of Guangdong Province, China, under Grant 2021B0101420006.

\bibliographystyle{cas-model2-names}

\bibliography{refs}

\begin{thebibliography}{136}
\expandafter\ifx\csname natexlab\endcsname\relax\def\natexlab#1{#1}\fi
\providecommand{\url}[1]{\texttt{#1}}
\providecommand{\href}[2]{#2}
\providecommand{\path}[1]{#1}
\providecommand{\DOIprefix}{doi:}
\providecommand{\ArXivprefix}{arXiv:}
\providecommand{\URLprefix}{URL: }
\providecommand{\Pubmedprefix}{pmid:}
\providecommand{\doi}[1]{\href{http://dx.doi.org/#1}{\path{#1}}}
\providecommand{\Pubmed}[1]{\href{pmid:#1}{\path{#1}}}
\providecommand{\bibinfo}[2]{#2}
\ifx\xfnm\relax \def\xfnm[#1]{\unskip,\space#1}\fi
\bibitem[{Abouhawwash and Alessio(2022)}]{abouhawwash2022evolutionary}
\bibinfo{author}{Abouhawwash, M.}, \bibinfo{author}{Alessio, A.}, \bibinfo{year}{2022}.
\newblock \bibinfo{title}{Evolutionary optimization of multiple machine-learned objectives for {PET} image reconstruction}.
\newblock \bibinfo{journal}{IEEE Transactions on Radiation and Plasma Medical Sciences} \bibinfo{volume}{7}, \bibinfo{pages}{273--283}.
\bibitem[{Amirrashedi et~al.(2021)Amirrashedi, Sarkar, Ghadiri, Ghafarian and Ay}]{2021Standard}
\bibinfo{author}{Amirrashedi, M.}, \bibinfo{author}{Sarkar, S.}, \bibinfo{author}{Ghadiri, H.}, \bibinfo{author}{Ghafarian, P.}, \bibinfo{author}{Ay, M.}, \bibinfo{year}{2021}.
\newblock \bibinfo{title}{Standard-dose {PET} reconstruction from low-dose preclinical images using an adopted all convolutional {U-N}et}.
\newblock \bibinfo{journal}{Biomedical Applications in Molecular, Structural, and Functional Imaging} \bibinfo{volume}{11600}, \bibinfo{pages}{834--848}.
\bibitem[{An et~al.(2016)An, Zhang, Adeli, Wang, Ma, Shi, Lalush, Lin and Shen}]{2016Multi}
\bibinfo{author}{An, L.}, \bibinfo{author}{Zhang, P.}, \bibinfo{author}{Adeli, E.}, \bibinfo{author}{Wang, Y.}, \bibinfo{author}{Ma, G.}, \bibinfo{author}{Shi, F.}, \bibinfo{author}{Lalush, D.S.}, \bibinfo{author}{Lin, W.}, \bibinfo{author}{Shen, D.}, \bibinfo{year}{2016}.
\newblock \bibinfo{title}{Multi-level canonical correlation analysis for standard-dose {PET} image estimation}.
\newblock \bibinfo{journal}{IEEE Transactions on Image Processing} \bibinfo{volume}{25}, \bibinfo{pages}{3303--3315}.
\bibitem[{Arabi and Zaidi(2018)}]{arabi2018improvement}
\bibinfo{author}{Arabi, H.}, \bibinfo{author}{Zaidi, H.}, \bibinfo{year}{2018}.
\newblock \bibinfo{title}{Improvement of image quality in {PET} using post-reconstruction hybrid spatial-frequency domain filtering}.
\newblock \bibinfo{journal}{Physics in Medicine \& Biology} \bibinfo{volume}{63}, \bibinfo{pages}{215010}.
\bibitem[{Avril and Weber(2005)}]{2005Monitoring}
\bibinfo{author}{Avril, N.E.}, \bibinfo{author}{Weber, W.A.}, \bibinfo{year}{2005}.
\newblock \bibinfo{title}{Monitoring response to treatment in patients utilizing {PET}.}
\newblock \bibinfo{journal}{Radiologic Clinics} \bibinfo{volume}{43}, \bibinfo{pages}{189--204}.
\bibitem[{Ba et~al.(2016)Ba, Kiros and Hinton}]{2016Layer}
\bibinfo{author}{Ba, J.L.}, \bibinfo{author}{Kiros, J.R.}, \bibinfo{author}{Hinton, G.E.}, \bibinfo{year}{2016}.
\newblock \bibinfo{title}{Layer normalization}.
\newblock \bibinfo{journal}{arXiv preprint arXiv:1607.06450} .
\bibitem[{Bailey et~al.(2006)Bailey, Maisey, Townsend and Valk}]{2006Positron}
\bibinfo{author}{Bailey, D.L.}, \bibinfo{author}{Maisey, M.N.}, \bibinfo{author}{Townsend, D.W.}, \bibinfo{author}{Valk, P.E.}, \bibinfo{year}{2006}.
\newblock \bibinfo{title}{Positron {E}mission {T}omography : {B}asic sciences}.
\newblock \bibinfo{journal}{Journal of Neuroradiology} \bibinfo{volume}{33}, \bibinfo{pages}{265--265}.
\bibitem[{Barbosa et~al.(2020)Barbosa, Queiroz, Nunes, Costa, Zaniboni, Marin, Cerri and Buchpiguel}]{de2020nonprostatic}
\bibinfo{author}{Barbosa, F.}, \bibinfo{author}{Queiroz, M.A.}, \bibinfo{author}{Nunes, R.F.}, \bibinfo{author}{Costa, L.B.}, \bibinfo{author}{Zaniboni, E.C.}, \bibinfo{author}{Marin, J.G.}, \bibinfo{author}{Cerri, G.G.}, \bibinfo{author}{Buchpiguel, C.A.}, \bibinfo{year}{2020}.
\newblock \bibinfo{title}{Nonprostatic diseases on {PSMA} {PET} imaging: a spectrum of benign and malignant findings}.
\newblock \bibinfo{journal}{Cancer Imaging} \bibinfo{volume}{20}, \bibinfo{pages}{1--23}.
\bibitem[{Bhavana and Krishnappa(2015)}]{bhavana2015multi}
\bibinfo{author}{Bhavana, V.}, \bibinfo{author}{Krishnappa, H.}, \bibinfo{year}{2015}.
\newblock \bibinfo{title}{Multi-modality medical image fusion using discrete wavelet transform}.
\newblock \bibinfo{journal}{Procedia Computer Science} \bibinfo{volume}{70}, \bibinfo{pages}{625--631}.
\bibitem[{Breiman(2001)}]{breiman2001random}
\bibinfo{author}{Breiman, L.}, \bibinfo{year}{2001}.
\newblock \bibinfo{title}{Random forests}.
\newblock \bibinfo{journal}{Machine Learning} \bibinfo{volume}{45}, \bibinfo{pages}{5--32}.
\bibitem[{Brunet et~al.(2011)Brunet, Vrscay and Wang}]{brunet2011mathematical}
\bibinfo{author}{Brunet, D.}, \bibinfo{author}{Vrscay, E.}, \bibinfo{author}{Wang, Z.}, \bibinfo{year}{2011}.
\newblock \bibinfo{title}{On the mathematical properties of the structural similarity index}.
\newblock \bibinfo{journal}{IEEE Transactions on Image Processing} \bibinfo{volume}{21}, \bibinfo{pages}{1488--1499}.
\bibitem[{Buades et~al.(2005)Buades, Coll and Morel}]{buades2005non}
\bibinfo{author}{Buades, A.}, \bibinfo{author}{Coll, B.}, \bibinfo{author}{Morel, J.}, \bibinfo{year}{2005}.
\newblock \bibinfo{title}{A non-local algorithm for image denoising}.
\newblock \bibinfo{journal}{2005 IEEE computer society conference on Computer Vision and Pattern Recognition} \bibinfo{volume}{2}, \bibinfo{pages}{60--65}.
\bibitem[{Buchbender et~al.(2012)Buchbender, Heusner, Lauenstein, Bockisch and Antoch}]{2012Oncologic}
\bibinfo{author}{Buchbender, C.}, \bibinfo{author}{Heusner, T.A.}, \bibinfo{author}{Lauenstein, T.C.}, \bibinfo{author}{Bockisch, A.}, \bibinfo{author}{Antoch, G.}, \bibinfo{year}{2012}.
\newblock \bibinfo{title}{Oncologic {PET/MRI}, part 1: tumors of the brain, head and neck, chest, abdomen, and pelvis.}
\newblock \bibinfo{journal}{Journal of Nuclear Medicine} \bibinfo{volume}{53}, \bibinfo{pages}{928--938}.
\bibitem[{Byrd et~al.(2023)Byrd, Dasari, Jansen and Kinahan}]{byrd2023impact}
\bibinfo{author}{Byrd, D.}, \bibinfo{author}{Dasari, P.}, \bibinfo{author}{Jansen, F.}, \bibinfo{author}{Kinahan, P.}, \bibinfo{year}{2023}.
\newblock \bibinfo{title}{Impact of subsets and iterations on {PET} image quality in {TOF-OSEM} reconstruction} , \bibinfo{pages}{1--1}.
\bibitem[{Carion et~al.(2020)Carion, Massa, Synnaeve, Usunier, Kirillov and Zagoruyko}]{carion2020end}
\bibinfo{author}{Carion, N.}, \bibinfo{author}{Massa, F.}, \bibinfo{author}{Synnaeve, G.}, \bibinfo{author}{Usunier, N.}, \bibinfo{author}{Kirillov, A.}, \bibinfo{author}{Zagoruyko, S.}, \bibinfo{year}{2020}.
\newblock \bibinfo{title}{End-to-end object detection with transformers}.
\newblock \bibinfo{journal}{European Conference on Computer Vision} , \bibinfo{pages}{213--229}.
\bibitem[{Chan et~al.(2009)Chan, Meikle, Fulton, Tian, Cai and Feng}]{chan2009non}
\bibinfo{author}{Chan, C.}, \bibinfo{author}{Meikle, S.}, \bibinfo{author}{Fulton, R.}, \bibinfo{author}{Tian, G.}, \bibinfo{author}{Cai, W.}, \bibinfo{author}{Feng, D.}, \bibinfo{year}{2009}.
\newblock \bibinfo{title}{A non-local post-filtering algorithm for {PET} incorporating anatomical knowledge}.
\newblock \bibinfo{journal}{2009 IEEE Nuclear Science Symposium Conference Record} , \bibinfo{pages}{2728--2732}.
\bibitem[{Chen(2007a)}]{2007Clinical}
\bibinfo{author}{Chen, W.}, \bibinfo{year}{2007}a.
\newblock \bibinfo{title}{Clinical applications of {PET} in brain tumors}.
\newblock \bibinfo{journal}{Journal of Nuclear Medicine} \bibinfo{volume}{48}, \bibinfo{pages}{1468--1481}.
\bibitem[{Chen(2007b)}]{chen2007clinical}
\bibinfo{author}{Chen, W.}, \bibinfo{year}{2007}b.
\newblock \bibinfo{title}{Clinical applications of {PET} in brain tumors}.
\newblock \bibinfo{journal}{Journal of Nuclear Medicine} \bibinfo{volume}{48}, \bibinfo{pages}{1468--1481}.
\bibitem[{Chen et~al.(2018)Chen, Badrinarayanan, Lee and Rabinovich}]{chen2018gradnorm}
\bibinfo{author}{Chen, Z.}, \bibinfo{author}{Badrinarayanan, V.}, \bibinfo{author}{Lee, C.}, \bibinfo{author}{Rabinovich, A.}, \bibinfo{year}{2018}.
\newblock \bibinfo{title}{Gradnorm: Gradient normalization for adaptive loss balancing in deep multitask networks}.
\newblock \bibinfo{journal}{International Conference on Machine Learning} , \bibinfo{pages}{794--803}.
\bibitem[{Cho et~al.(2010)Cho, Chun, Reese, Fakhri, Zhu, Ouyang, Catana and Guerin}]{cho2010compensation}
\bibinfo{author}{Cho, S.}, \bibinfo{author}{Chun, S.Y.}, \bibinfo{author}{Reese, T.}, \bibinfo{author}{Fakhri, G.}, \bibinfo{author}{Zhu, X.}, \bibinfo{author}{Ouyang, J.}, \bibinfo{author}{Catana, C.}, \bibinfo{author}{Guerin, B.}, \bibinfo{year}{2010}.
\newblock \bibinfo{title}{Compensation for nonrigid motion using {B}-spline image registration in simultaneous {MR-PET}}.
\newblock \bibinfo{journal}{International Society for Magnetic Resonance in Medicine} .
\bibitem[{{\c{C}}i{\c{c}}ek et~al.(2016){\c{C}}i{\c{c}}ek, Abdulkadir, Lienkamp, Brox and Ronneberger}]{cciccek20163d}
\bibinfo{author}{{\c{C}}i{\c{c}}ek, {\"O}.}, \bibinfo{author}{Abdulkadir, A.}, \bibinfo{author}{Lienkamp, S.}, \bibinfo{author}{Brox, T.}, \bibinfo{author}{Ronneberger, O.}, \bibinfo{year}{2016}.
\newblock \bibinfo{title}{3{D} {U-N}et: {L}earning dense volumetric segmentation from sparse annotation}.
\newblock \bibinfo{journal}{International Conference on Medical Image Computing and Computer-Assisted Intervention} , \bibinfo{pages}{424--432}.
\bibitem[{Cui et~al.(2019)Cui, Gong, Guo, Wu, Meng, Kim, Zheng, Wu, Fu, Xu et~al.}]{cui2019pet}
\bibinfo{author}{Cui, J.}, \bibinfo{author}{Gong, K.}, \bibinfo{author}{Guo, N.}, \bibinfo{author}{Wu, C.}, \bibinfo{author}{Meng, X.}, \bibinfo{author}{Kim, K.}, \bibinfo{author}{Zheng, K.}, \bibinfo{author}{Wu, Z.}, \bibinfo{author}{Fu, L.}, \bibinfo{author}{Xu, B.}, et~al., \bibinfo{year}{2019}.
\newblock \bibinfo{title}{P{ET} image denoising using unsupervised deep learning}.
\newblock \bibinfo{journal}{European journal of nuclear medicine and molecular imaging} \bibinfo{volume}{46}, \bibinfo{pages}{2780--2789}.
\bibitem[{Dabov et~al.(2006)Dabov, Foi, Katkovnik and Egiazarian}]{dabov2006image}
\bibinfo{author}{Dabov, K.}, \bibinfo{author}{Foi, A.}, \bibinfo{author}{Katkovnik, V.}, \bibinfo{author}{Egiazarian, K.}, \bibinfo{year}{2006}.
\newblock \bibinfo{title}{Image denoising with block-matching and 3{D} filtering} \bibinfo{volume}{6064}, \bibinfo{pages}{354--365}.
\bibitem[{Daerr et~al.(2016)Daerr, Brendel, Zach, Mille and Rominger}]{2016Evaluation}
\bibinfo{author}{Daerr, S.}, \bibinfo{author}{Brendel, M.}, \bibinfo{author}{Zach, C.}, \bibinfo{author}{Mille, E.}, \bibinfo{author}{Rominger, A.}, \bibinfo{year}{2016}.
\newblock \bibinfo{title}{Evaluation of early-phase [18{F}]-florbetaben {PET} acquisition in clinical routine cases}.
\newblock \bibinfo{journal}{Neuroimage Clinical} \bibinfo{volume}{14}.
\bibitem[{Daubechies(1990)}]{daubechies1990wavelet}
\bibinfo{author}{Daubechies, I.}, \bibinfo{year}{1990}.
\newblock \bibinfo{title}{The wavelet transform, time-frequency localization and signal analysis}.
\newblock \bibinfo{journal}{IEEE transactions on information theory} \bibinfo{volume}{36}, \bibinfo{pages}{961--1005}.
\bibitem[{Daubechies(2009)}]{daubechies2009wavelet}
\bibinfo{author}{Daubechies, I.}, \bibinfo{year}{2009}.
\newblock \bibinfo{title}{The wavelet transform, time-frequency localization and signal analysis} .
\bibitem[{Decazes et~al.(2021)Decazes, Hinault, Veresezan, Thureau, Gouel and Vera}]{decazes2021trimodality}
\bibinfo{author}{Decazes, P.}, \bibinfo{author}{Hinault, P.}, \bibinfo{author}{Veresezan, O.}, \bibinfo{author}{Thureau, S.}, \bibinfo{author}{Gouel, P.}, \bibinfo{author}{Vera, P.}, \bibinfo{year}{2021}.
\newblock \bibinfo{title}{Trimodality {PET/CT/MRI} and radiotherapy: {A} mini-review}.
\newblock \bibinfo{journal}{Frontiers in Oncology} \bibinfo{volume}{10}, \bibinfo{pages}{3392}.
\bibitem[{Diaz-Pinto et~al.(2019)Diaz-Pinto, Colomer, Morales, Xu and Frangi}]{diaz2019retinal}
\bibinfo{author}{Diaz-Pinto, A.}, \bibinfo{author}{Colomer, A.}, \bibinfo{author}{Morales, S.}, \bibinfo{author}{Xu, Y.}, \bibinfo{author}{Frangi, A.}, \bibinfo{year}{2019}.
\newblock \bibinfo{title}{Retinal image synthesis and semi-supervised learning for glaucoma assessment}.
\newblock \bibinfo{journal}{IEEE Transactions on Medical Imaging} \bibinfo{volume}{38}, \bibinfo{pages}{2211--2218}.
\bibitem[{Dimitrakopoulou-Strauss et~al.(2021)Dimitrakopoulou-Strauss, Pan and Sachpekidis}]{dimitrakopoulou2021kinetic}
\bibinfo{author}{Dimitrakopoulou-Strauss, A.}, \bibinfo{author}{Pan, L.}, \bibinfo{author}{Sachpekidis, C.}, \bibinfo{year}{2021}.
\newblock \bibinfo{title}{Kinetic modeling and parametric imaging with dynamic {PET} for oncological applications: {G}eneral considerations, current clinical applications, and future perspectives}.
\newblock \bibinfo{journal}{European Journal of Nuclear Medicine and Molecular Imaging} \bibinfo{volume}{48}, \bibinfo{pages}{21--39}.
\bibitem[{Dutta et~al.(2013)Dutta, Leahy and Li}]{dutta2013non}
\bibinfo{author}{Dutta, J.}, \bibinfo{author}{Leahy, R.}, \bibinfo{author}{Li, Q.}, \bibinfo{year}{2013}.
\newblock \bibinfo{title}{Non-local means denoising of dynamic {PET} images}.
\newblock \bibinfo{journal}{PloS One} \bibinfo{volume}{8}, \bibinfo{pages}{e81390}.
\bibitem[{E.~Samadi et~al.(2022)E.~Samadi, Kiefer, Fritsch, Bickenbach and Schuppert}]{e2022training}
\bibinfo{author}{E.~Samadi, M.}, \bibinfo{author}{Kiefer, S.}, \bibinfo{author}{Fritsch, S.J.}, \bibinfo{author}{Bickenbach, J.}, \bibinfo{author}{Schuppert, A.}, \bibinfo{year}{2022}.
\newblock \bibinfo{title}{A training strategy for hybrid models to break the curse of dimensionality}.
\newblock \bibinfo{journal}{Plos One} \bibinfo{volume}{17}, \bibinfo{pages}{e0274569}.
\bibitem[{Fahey(2002)}]{fahey2002data}
\bibinfo{author}{Fahey, F.}, \bibinfo{year}{2002}.
\newblock \bibinfo{title}{Data acquisition in {PET} imaging}.
\newblock \bibinfo{journal}{Journal of Nuclear Medicine Technology} \bibinfo{volume}{30}, \bibinfo{pages}{39--49}.
\bibitem[{Feng et~al.(2022)Feng, Qin, Chai, Zeng, Wang, Meng and Wang}]{feng2022mri}
\bibinfo{author}{Feng, E.}, \bibinfo{author}{Qin, P.}, \bibinfo{author}{Chai, R.}, \bibinfo{author}{Zeng, J.}, \bibinfo{author}{Wang, Q.}, \bibinfo{author}{Meng, Y.}, \bibinfo{author}{Wang, P.}, \bibinfo{year}{2022}.
\newblock \bibinfo{title}{M{RI} generated from {CT} for acute ischemic stroke combining radiomics and generative adversarial networks}.
\newblock \bibinfo{journal}{IEEE Journal of Biomedical and Health Informatics} .
\bibitem[{Feng and Liu(2020)}]{feng2020rethinking}
\bibinfo{author}{Feng, Q.}, \bibinfo{author}{Liu, H.}, \bibinfo{year}{2020}.
\newblock \bibinfo{title}{Rethinking {PET} image reconstruction: {U}ltra-low-dose, sinogram and deep learning}.
\newblock \bibinfo{journal}{International Conference on Medical Image Computing and Computer-Assisted Intervention} , \bibinfo{pages}{783--792}.
\bibitem[{Fortunato(2010)}]{Fortunato2010}
\bibinfo{author}{Fortunato, S.}, \bibinfo{year}{2010}.
\newblock \bibinfo{title}{Community detection in graphs}.
\newblock \bibinfo{journal}{Phys. Rep.-Rev. Sec. Phys. Lett.} \bibinfo{volume}{486}, \bibinfo{pages}{75--174}.
\bibitem[{Geng et~al.(2024)Geng, Tan, Wang, Jia, Zhang and Yan}]{geng2024stcnet}
\bibinfo{author}{Geng, P.}, \bibinfo{author}{Tan, Z.}, \bibinfo{author}{Wang, Y.}, \bibinfo{author}{Jia, W.}, \bibinfo{author}{Zhang, Y.}, \bibinfo{author}{Yan, H.}, \bibinfo{year}{2024}.
\newblock \bibinfo{title}{{STCNet}: Alternating {CNN} and improved transformer network for {COVID}-19 {CT} image segmentation}.
\newblock \bibinfo{journal}{Biomedical Signal Processing and Control} \bibinfo{volume}{93}, \bibinfo{pages}{106205}.
\bibitem[{Gigengack et~al.(2012)Gigengack, Ruthotto, Burger, Wolters, Jiang and Schafers}]{2012Motion}
\bibinfo{author}{Gigengack, F.}, \bibinfo{author}{Ruthotto, L.}, \bibinfo{author}{Burger, M.}, \bibinfo{author}{Wolters, C.H.}, \bibinfo{author}{Jiang, X.}, \bibinfo{author}{Schafers, K.P.}, \bibinfo{year}{2012}.
\newblock \bibinfo{title}{Motion correction in dual gated cardiac {PET} using mass-preserving image registration.}
\newblock \bibinfo{journal}{IEEE Transactions on Medical Imaging} \bibinfo{volume}{31}, \bibinfo{pages}{698--712}.
\bibitem[{Goodfellow et~al.(2014)Goodfellow, Pouget-Abadie, Mirza, Xu, Warde-Farley, Ozair, Courville and Bengio}]{goodfellow2014generative}
\bibinfo{author}{Goodfellow, I.}, \bibinfo{author}{Pouget-Abadie, J.}, \bibinfo{author}{Mirza, M.}, \bibinfo{author}{Xu, B.}, \bibinfo{author}{Warde-Farley, D.}, \bibinfo{author}{Ozair, S.}, \bibinfo{author}{Courville, A.}, \bibinfo{author}{Bengio, Y.}, \bibinfo{year}{2014}.
\newblock \bibinfo{title}{Generative adversarial nets}.
\newblock \bibinfo{journal}{Advances in Neural Information Processing Systems} \bibinfo{volume}{27}, \bibinfo{pages}{100--109}.
\bibitem[{Gu et~al.(2021)Gu, Yang, Ye and Yang}]{gu2021cyclegan}
\bibinfo{author}{Gu, J.}, \bibinfo{author}{Yang, T.}, \bibinfo{author}{Ye, J.}, \bibinfo{author}{Yang, D.}, \bibinfo{year}{2021}.
\newblock \bibinfo{title}{Cycle{GAN} denoising of extreme low-dose cardiac {CT} using wavelet-assisted noise disentanglement}.
\newblock \bibinfo{journal}{Medical Image Analysis} \bibinfo{volume}{74}, \bibinfo{pages}{102209}.
\bibitem[{H{\"a}ggstr{\"o}m et~al.(2019)H{\"a}ggstr{\"o}m, Schmidtlein, Campanella and Fuchs}]{haggstrom2019deeppet}
\bibinfo{author}{H{\"a}ggstr{\"o}m, I.}, \bibinfo{author}{Schmidtlein, C.}, \bibinfo{author}{Campanella, G.}, \bibinfo{author}{Fuchs, T.}, \bibinfo{year}{2019}.
\newblock \bibinfo{title}{Deep{PET}: {A} deep encoder--decoder network for directly solving the {PET} image reconstruction inverse problem}.
\newblock \bibinfo{journal}{Medical Image Analysis} \bibinfo{volume}{54}, \bibinfo{pages}{253--262}.
\bibitem[{Hamill et~al.(2003)Hamill, Michel and Kinahan}]{hamill2003fast}
\bibinfo{author}{Hamill, J.}, \bibinfo{author}{Michel, C.}, \bibinfo{author}{Kinahan, P.}, \bibinfo{year}{2003}.
\newblock \bibinfo{title}{Fast {PET} {EM} reconstruction from linograms}.
\newblock \bibinfo{journal}{IEEE Transactions on Nuclear Science} \bibinfo{volume}{50}, \bibinfo{pages}{1630--1635}.
\bibitem[{H{\"a}ndel(2018)}]{handel2018understanding}
\bibinfo{author}{H{\"a}ndel, P.}, \bibinfo{year}{2018}.
\newblock \bibinfo{title}{Understanding normalized mean squared error in power amplifier linearization}.
\newblock \bibinfo{journal}{IEEE Microwave and Wireless Components Letters} \bibinfo{volume}{28}, \bibinfo{pages}{1047--1049}.
\bibitem[{Hofheinz et~al.(2011)Hofheinz, Langner, Beuthien-Baumann, Oehme, Steinbach, Kotzerke and Hoff}]{hofheinz2011suitability}
\bibinfo{author}{Hofheinz, F.}, \bibinfo{author}{Langner, J.}, \bibinfo{author}{Beuthien-Baumann, B.}, \bibinfo{author}{Oehme, L.}, \bibinfo{author}{Steinbach, J.}, \bibinfo{author}{Kotzerke, J.}, \bibinfo{author}{Hoff, J.}, \bibinfo{year}{2011}.
\newblock \bibinfo{title}{Suitability of bilateral filtering for edge-preserving noise reduction in {PET}}.
\newblock \bibinfo{journal}{EJNMMI Research} \bibinfo{volume}{1}, \bibinfo{pages}{1--9}.
\bibitem[{Hong et~al.(2023)Hong, Lin, Li, Li, Yao, Wu, Liu and Tian}]{hong2023distance}
\bibinfo{author}{Hong, Q.}, \bibinfo{author}{Lin, L.}, \bibinfo{author}{Li, Z.}, \bibinfo{author}{Li, Q.}, \bibinfo{author}{Yao, J.}, \bibinfo{author}{Wu, Q.}, \bibinfo{author}{Liu, K.}, \bibinfo{author}{Tian, J.}, \bibinfo{year}{2023}.
\newblock \bibinfo{title}{A distance transformation deep forest framework with hybrid-feature fusion for {CRX} image classification}.
\newblock \bibinfo{journal}{IEEE Transactions on Neural Networks and Learning Systems} .
\bibitem[{Hu et~al.(2017)Hu, Zhang, Zhang, Liao and Wang}]{2017Low}
\bibinfo{author}{Hu, C.}, \bibinfo{author}{Zhang, Y.}, \bibinfo{author}{Zhang, W.}, \bibinfo{author}{Liao, P.}, \bibinfo{author}{Wang, G.}, \bibinfo{year}{2017}.
\newblock \bibinfo{title}{Low-dose {CT} via convolutional neural network}.
\newblock \bibinfo{journal}{Biomedical Optics Express} \bibinfo{volume}{8}, \bibinfo{pages}{679}.
\bibitem[{Hu et~al.(2020)Hu, Yu, Chen and Wang}]{hu2020medical}
\bibinfo{author}{Hu, S.}, \bibinfo{author}{Yu, W.}, \bibinfo{author}{Chen, Z.}, \bibinfo{author}{Wang, S.}, \bibinfo{year}{2020}.
\newblock \bibinfo{title}{Medical image reconstruction using generative adversarial network for alzheimer disease assessment with class-imbalance problem}.
\newblock \bibinfo{journal}{2020 IEEE 6th International Conference on Computer and Communications (ICCC)} , \bibinfo{pages}{1323--1327}.
\bibitem[{Huang et~al.(2022)Huang, Guan, Chen, Zhu, Chen and Yu}]{huang2022deep}
\bibinfo{author}{Huang, J.}, \bibinfo{author}{Guan, L.}, \bibinfo{author}{Chen, Y.}, \bibinfo{author}{Zhu, S.}, \bibinfo{author}{Chen, L.}, \bibinfo{author}{Yu, J.}, \bibinfo{year}{2022}.
\newblock \bibinfo{title}{A deep learning scheme for transient stability assessment in power system with a hierarchical dynamic graph pooling method}.
\newblock \bibinfo{journal}{International Journal of Electrical Power \& Energy Systems} \bibinfo{volume}{141}, \bibinfo{pages}{108044}.
\bibitem[{Hullermeier and Rifqi(2009)}]{HullermeierRifqi2009}
\bibinfo{author}{Hullermeier, E.}, \bibinfo{author}{Rifqi, M.}, \bibinfo{year}{2009}.
\newblock \bibinfo{title}{A fuzzy variant of the rand index for comparing clustering structures}, in: \bibinfo{booktitle}{in Proc. IFSA/EUSFLAT Conf.}, pp. \bibinfo{pages}{1294--1298}.
\bibitem[{Ioffe and Szegedy(2015)}]{2015Batch}
\bibinfo{author}{Ioffe, S.}, \bibinfo{author}{Szegedy, C.}, \bibinfo{year}{2015}.
\newblock \bibinfo{title}{Batch normalization: {A}ccelerating deep network training by reducing internal covariate shift}.
\newblock \bibinfo{journal}{International Conference on Machine Learning} , \bibinfo{pages}{448--456}.
\bibitem[{Isola et~al.(2017)Isola, Zhu, Zhou and Efros}]{isola2017image}
\bibinfo{author}{Isola, P.}, \bibinfo{author}{Zhu, J.}, \bibinfo{author}{Zhou, T.}, \bibinfo{author}{Efros, A.}, \bibinfo{year}{2017}.
\newblock \bibinfo{title}{Image-to-image translation with conditional adversarial networks}.
\newblock \bibinfo{journal}{Proceedings of the IEEE conference on computer vision and pattern recognition} , \bibinfo{pages}{1125--1134}.
\bibitem[{Jia et~al.(2010)Jia, Wu, Wang and Shen}]{jia2010absorb}
\bibinfo{author}{Jia, H.}, \bibinfo{author}{Wu, G.}, \bibinfo{author}{Wang, Q.}, \bibinfo{author}{Shen, D.}, \bibinfo{year}{2010}.
\newblock \bibinfo{title}{A{BSORB}: Atlas building by self-organized registration and bundling}.
\newblock \bibinfo{journal}{NeuroImage} \bibinfo{volume}{51}, \bibinfo{pages}{1057--1070}.
\bibitem[{Jia et~al.(2012)Jia, Yap and Shen}]{jia2012iterative}
\bibinfo{author}{Jia, H.}, \bibinfo{author}{Yap, P.}, \bibinfo{author}{Shen, D.}, \bibinfo{year}{2012}.
\newblock \bibinfo{title}{Iterative multi-atlas-based multi-image segmentation with tree-based registration}.
\newblock \bibinfo{journal}{NeuroImage} \bibinfo{volume}{59}, \bibinfo{pages}{422--430}.
\bibitem[{Jiang et~al.(2022)Jiang, Pan, Cui and Shen}]{jiang2022reconstruction}
\bibinfo{author}{Jiang, C.}, \bibinfo{author}{Pan, Y.}, \bibinfo{author}{Cui, Z.}, \bibinfo{author}{Shen, D.}, \bibinfo{year}{2022}.
\newblock \bibinfo{title}{Reconstruction of standard-dose {PET} from low-dose {PET} via dual-frequency supervision and global aggregation module}.
\newblock \bibinfo{journal}{2022 IEEE 19th International Symposium on Biomedical Imaging (ISBI)} , \bibinfo{pages}{1--5}.
\bibitem[{Jiang et~al.(2023)Jiang, Pan and Shen}]{jiang2023tridornet}
\bibinfo{author}{Jiang, C.}, \bibinfo{author}{Pan, Y.}, \bibinfo{author}{Shen, D.}, \bibinfo{year}{2023}.
\newblock \bibinfo{title}{Tridornet: Reconstruction of standard-dose {PET} from low-dose {PET} in triple (projection, image, and frequency) domains} , \bibinfo{pages}{1--5}.
\bibitem[{Jiang et~al.(2021)Jiang, Dai, Wu and Loy}]{jiang2021focal}
\bibinfo{author}{Jiang, L.}, \bibinfo{author}{Dai, B.}, \bibinfo{author}{Wu, W.}, \bibinfo{author}{Loy, C.C.}, \bibinfo{year}{2021}.
\newblock \bibinfo{title}{Focal frequency loss for image reconstruction and synthesis}.
\newblock \bibinfo{journal}{Proceedings of the IEEE/CVF International Conference on Computer Vision} , \bibinfo{pages}{13919--13929}.
\bibitem[{Kamnitsas et~al.(2016)Kamnitsas, Ledig, Newcombe, Simpson, Kane, Menon, Rueckert and Glocker}]{2016Efficient}
\bibinfo{author}{Kamnitsas, K.}, \bibinfo{author}{Ledig, C.}, \bibinfo{author}{Newcombe, V.}, \bibinfo{author}{Simpson, J.P.}, \bibinfo{author}{Kane, A.D.}, \bibinfo{author}{Menon, D.K.}, \bibinfo{author}{Rueckert, D.}, \bibinfo{author}{Glocker, B.}, \bibinfo{year}{2016}.
\newblock \bibinfo{title}{Efficient multi-scale 3{D} {CNN} with fully connected {CRF} for accurate brain lesion segmentation}.
\newblock \bibinfo{journal}{Medical Image Analysis} \bibinfo{volume}{36}, \bibinfo{pages}{61}.
\bibitem[{Kamran et~al.(2021)Kamran, Hossain, Tavakkoli, Zuckerbrod and Baker}]{kamran2021vtgan}
\bibinfo{author}{Kamran, S.}, \bibinfo{author}{Hossain, K.}, \bibinfo{author}{Tavakkoli, A.}, \bibinfo{author}{Zuckerbrod, S.}, \bibinfo{author}{Baker, S.}, \bibinfo{year}{2021}.
\newblock \bibinfo{title}{Vt{GAN}: Semi-supervised retinal image synthesis and disease prediction using vision transformers}.
\newblock \bibinfo{journal}{Proceedings of the IEEE/CVF International Conference on Computer Vision} , \bibinfo{pages}{3235--3245}.
\bibitem[{Kang et~al.(2015)Kang, Gao, Shi, Lalush, Lin and Shen}]{2015Prediction}
\bibinfo{author}{Kang, J.}, \bibinfo{author}{Gao, Y.}, \bibinfo{author}{Shi, F.}, \bibinfo{author}{Lalush, D.S.}, \bibinfo{author}{Lin, W.}, \bibinfo{author}{Shen, D.}, \bibinfo{year}{2015}.
\newblock \bibinfo{title}{Prediction of standard-dose brain {PET} image by using {MRI} and low-dose brain [18{F}]{FDG} {PET} images}.
\newblock \bibinfo{journal}{Medical Physics} \bibinfo{volume}{42}, \bibinfo{pages}{5301--5309}.
\bibitem[{Katsevich(2004)}]{katsevich2004improved}
\bibinfo{author}{Katsevich, A.}, \bibinfo{year}{2004}.
\newblock \bibinfo{title}{An improved exact filtered backprojection algorithm for spiral computed tomography}.
\newblock \bibinfo{journal}{Advances in Applied Mathematics} \bibinfo{volume}{32}, \bibinfo{pages}{681--697}.
\bibitem[{Kawahara et~al.(2017)Kawahara, Brown, Miller, Booth and Hamarneh}]{2017BrainNetCNN}
\bibinfo{author}{Kawahara, J.}, \bibinfo{author}{Brown, C.J.}, \bibinfo{author}{Miller, S.P.}, \bibinfo{author}{Booth, B.G.}, \bibinfo{author}{Hamarneh, G.}, \bibinfo{year}{2017}.
\newblock \bibinfo{title}{Brain{NetCNN}: {C}onvolutional neural networks for brain networks; towards predicting neurodevelopment}.
\newblock \bibinfo{journal}{NeuroImage} \bibinfo{volume}{146}, \bibinfo{pages}{1038--1049}.
\bibitem[{Kim et~al.(2018)Kim, Wu, Gong, Dutta, Kim, Son, Kim, F. and Li}]{2018Penalized}
\bibinfo{author}{Kim, K.}, \bibinfo{author}{Wu, D.}, \bibinfo{author}{Gong, K.}, \bibinfo{author}{Dutta, J.}, \bibinfo{author}{Kim, J.}, \bibinfo{author}{Son, Y.}, \bibinfo{author}{Kim, H.}, \bibinfo{author}{F., E.}, \bibinfo{author}{Li, Q.}, \bibinfo{year}{2018}.
\newblock \bibinfo{title}{Penalized {PET} reconstruction using deep learning prior and local linear fitting}.
\newblock \bibinfo{journal}{IEEE Transactions on Medical Imaging} \bibinfo{volume}{37}, \bibinfo{pages}{1478--1487}.
\bibitem[{Kleesiek et~al.(2016)Kleesiek, Urban, Hubert, Schwarz, Maier-Hein, Bendszus and Biller}]{2016Deep}
\bibinfo{author}{Kleesiek, J.}, \bibinfo{author}{Urban, G.}, \bibinfo{author}{Hubert, A.}, \bibinfo{author}{Schwarz, D.}, \bibinfo{author}{Maier-Hein, K.}, \bibinfo{author}{Bendszus, M.}, \bibinfo{author}{Biller, A.}, \bibinfo{year}{2016}.
\newblock \bibinfo{title}{Deep {MRI} brain extraction: {A} 3{D} convolutional neural network for skull stripping}.
\newblock \bibinfo{journal}{NeuroImage} \bibinfo{volume}{129}, \bibinfo{pages}{460--469}.
\bibitem[{Korhonen and You(2012)}]{korhonen2012peak}
\bibinfo{author}{Korhonen, J.}, \bibinfo{author}{You, J.}, \bibinfo{year}{2012}.
\newblock \bibinfo{title}{Peak signal-to-noise ratio revisited: {I}s simple beautiful?}
\newblock \bibinfo{journal}{2012 Fourth International Workshop on Quality of Multimedia Experience} , \bibinfo{pages}{37--38}.
\bibitem[{Kreisl et~al.(2020)Kreisl, Kim, Coughlin, Henter, Owen and Innis}]{kreisl2020pet}
\bibinfo{author}{Kreisl, W.C.}, \bibinfo{author}{Kim, M.}, \bibinfo{author}{Coughlin, J.M.}, \bibinfo{author}{Henter, I.D.}, \bibinfo{author}{Owen, D.R.}, \bibinfo{author}{Innis, R.B.}, \bibinfo{year}{2020}.
\newblock \bibinfo{title}{P{ET} imaging of neuroinflammation in neurological disorders}.
\newblock \bibinfo{journal}{The Lancet Neurology} \bibinfo{volume}{19}, \bibinfo{pages}{940--950}.
\bibitem[{Lehtinen et~al.(2018)Lehtinen, Munkberg, Hasselgren, Laine, Karras, Aittala and Aila}]{lehtinen2018noise2noise}
\bibinfo{author}{Lehtinen, J.}, \bibinfo{author}{Munkberg, J.}, \bibinfo{author}{Hasselgren, J.}, \bibinfo{author}{Laine, S.}, \bibinfo{author}{Karras, T.}, \bibinfo{author}{Aittala, M.}, \bibinfo{author}{Aila, T.}, \bibinfo{year}{2018}.
\newblock \bibinfo{title}{Noise2{N}oise: Learning image restoration without clean data}.
\newblock \bibinfo{journal}{arXiv preprint arXiv:1803.04189} .
\bibitem[{Lei et~al.(2020)Lei, Dong, Wang, Higgins and Yang}]{2020Estimating}
\bibinfo{author}{Lei, Y.}, \bibinfo{author}{Dong, X.}, \bibinfo{author}{Wang, T.}, \bibinfo{author}{Higgins, K.}, \bibinfo{author}{Yang, X.}, \bibinfo{year}{2020}.
\newblock \bibinfo{title}{Estimating standard-dose {PET} from low-dose {PET} with deep learning}.
\newblock \bibinfo{journal}{Image Processing} \bibinfo{volume}{113}, \bibinfo{pages}{73--82}.
\bibitem[{Li et~al.(2022)Li, Cui, Chen, Zeng, Wollenweber, Jansen, Jang, Kim, Gong and Li}]{li2022noise}
\bibinfo{author}{Li, Y.}, \bibinfo{author}{Cui, J.}, \bibinfo{author}{Chen, J.}, \bibinfo{author}{Zeng, G.}, \bibinfo{author}{Wollenweber, S.}, \bibinfo{author}{Jansen, F.}, \bibinfo{author}{Jang, S.}, \bibinfo{author}{Kim, K.}, \bibinfo{author}{Gong, K.}, \bibinfo{author}{Li, Q.}, \bibinfo{year}{2022}.
\newblock \bibinfo{title}{A noise-level-aware framework for {PET} image denoising}.
\newblock \bibinfo{journal}{arXiv preprint arXiv:2203.08034} .
\bibitem[{Li et~al.(2021)Li, Zhang, Shi, Miao and Jiang}]{li2021novel}
\bibinfo{author}{Li, Y.}, \bibinfo{author}{Zhang, K.}, \bibinfo{author}{Shi, W.}, \bibinfo{author}{Miao, Y.}, \bibinfo{author}{Jiang, Z.}, \bibinfo{year}{2021}.
\newblock \bibinfo{title}{A novel medical image denoising method based on conditional generative adversarial network}.
\newblock \bibinfo{journal}{Computational and Mathematical Methods in Medicine} \bibinfo{volume}{2021}, \bibinfo{pages}{1--11}.
\bibitem[{Li et~al.(2019)Li, Huang, Yu, Chi and Jin}]{li2019low}
\bibinfo{author}{Li, Z.}, \bibinfo{author}{Huang, J.}, \bibinfo{author}{Yu, L.}, \bibinfo{author}{Chi, Y.}, \bibinfo{author}{Jin, M.}, \bibinfo{year}{2019}.
\newblock \bibinfo{title}{Low-dose {CT} image denoising using cycle-consistent adversarial networks}.
\newblock \bibinfo{journal}{2019 IEEE Nuclear Science Symposium and Medical Imaging Conference (NSS/MIC)} , \bibinfo{pages}{1--3}.
\bibitem[{Liang et~al.(2022)Liang, Yang, Huang, Li, He, Hu, Chen, Xue, Cheng and Ni}]{liang2022sketch}
\bibinfo{author}{Liang, J.}, \bibinfo{author}{Yang, X.}, \bibinfo{author}{Huang, Y.}, \bibinfo{author}{Li, H.}, \bibinfo{author}{He, S.}, \bibinfo{author}{Hu, X.}, \bibinfo{author}{Chen, Z.}, \bibinfo{author}{Xue, W.}, \bibinfo{author}{Cheng, J.}, \bibinfo{author}{Ni, D.}, \bibinfo{year}{2022}.
\newblock \bibinfo{title}{Sketch guided and progressive growing {GAN} for realistic and editable ultrasound image synthesis}.
\newblock \bibinfo{journal}{Medical Image Analysis} \bibinfo{volume}{79}, \bibinfo{pages}{102461}.
\bibitem[{Liaw and Wiener(2002)}]{2002Classification}
\bibinfo{author}{Liaw, A.}, \bibinfo{author}{Wiener, M.}, \bibinfo{year}{2002}.
\newblock \bibinfo{title}{Classification and regression by randomforest}.
\newblock \bibinfo{journal}{R News} \bibinfo{volume}{2}, \bibinfo{pages}{18--22}.
\bibitem[{Liu et~al.(2018)Liu, Zhang, Zhang, Lin and Zuo}]{liu2018multi}
\bibinfo{author}{Liu, P.}, \bibinfo{author}{Zhang, H.}, \bibinfo{author}{Zhang, K.}, \bibinfo{author}{Lin, L.}, \bibinfo{author}{Zuo, W.}, \bibinfo{year}{2018}.
\newblock \bibinfo{title}{Multi-level wavelet-{CNN} for image restoration}.
\newblock \bibinfo{journal}{Proceedings of the IEEE conference on computer vision and pattern recognition workshops} , \bibinfo{pages}{773--782}.
\bibitem[{Liu et~al.(2021)Liu, Lin, Cao, Hu, Wei, Zhang, Lin and Guo}]{liu2021swin}
\bibinfo{author}{Liu, Z.}, \bibinfo{author}{Lin, Y.}, \bibinfo{author}{Cao, Y.}, \bibinfo{author}{Hu, H.}, \bibinfo{author}{Wei, Y.}, \bibinfo{author}{Zhang, Z.}, \bibinfo{author}{Lin, S.}, \bibinfo{author}{Guo, B.}, \bibinfo{year}{2021}.
\newblock \bibinfo{title}{Swin transformer: Hierarchical vision transformer using shifted windows}.
\newblock \bibinfo{journal}{Proceedings of the IEEE/CVF International Conference on Computer Vision} , \bibinfo{pages}{10012--10022}.
\bibitem[{Lu et~al.(2019)Lu, Onofrey, Lu, Shi, Ma, Liu and Liu}]{lu2019investigation}
\bibinfo{author}{Lu, W.}, \bibinfo{author}{Onofrey, J.}, \bibinfo{author}{Lu, Y.}, \bibinfo{author}{Shi, L.}, \bibinfo{author}{Ma, T.}, \bibinfo{author}{Liu, Y.}, \bibinfo{author}{Liu, C.}, \bibinfo{year}{2019}.
\newblock \bibinfo{title}{An investigation of quantitative accuracy for deep learning based denoising in oncological {PET}}.
\newblock \bibinfo{journal}{Physics in Medicine \& Biology} \bibinfo{volume}{64}, \bibinfo{pages}{165019}.
\bibitem[{Lu et~al.(2021)Lu, Li, Wang and Shen}]{lu2021two}
\bibinfo{author}{Lu, Z.}, \bibinfo{author}{Li, Z.}, \bibinfo{author}{Wang, J.}, \bibinfo{author}{Shen, D.}, \bibinfo{year}{2021}.
\newblock \bibinfo{title}{Two-stage self-supervised cycle-consistency network for reconstruction of thin-slice {MR} images}.
\newblock \bibinfo{journal}{arXiv preprint arXiv:2106.15395} .
\bibitem[{Luo et~al.(2021)Luo, Wang, Zu, Zhan, Wu, Zhou, Shen and Zhou}]{luo20213d}
\bibinfo{author}{Luo, Y.}, \bibinfo{author}{Wang, Y.}, \bibinfo{author}{Zu, C.}, \bibinfo{author}{Zhan, B.}, \bibinfo{author}{Wu, X.}, \bibinfo{author}{Zhou, J.}, \bibinfo{author}{Shen, D.}, \bibinfo{author}{Zhou, L.}, \bibinfo{year}{2021}.
\newblock \bibinfo{title}{3{D} transformer-{GAN} for high-quality {PET} reconstruction}.
\newblock \bibinfo{journal}{International Conference on Medical Image Computing and Computer-Assisted Intervention} , \bibinfo{pages}{276--285}.
\bibitem[{Luo et~al.(2022)Luo, Zhou, Zhan, Fei, Zhou, Wang and Shen}]{luo2022adaptive}
\bibinfo{author}{Luo, Y.}, \bibinfo{author}{Zhou, L.}, \bibinfo{author}{Zhan, B.}, \bibinfo{author}{Fei, Y.}, \bibinfo{author}{Zhou, J.}, \bibinfo{author}{Wang, Y.}, \bibinfo{author}{Shen, D.}, \bibinfo{year}{2022}.
\newblock \bibinfo{title}{Adaptive rectification based adversarial network with spectrum constraint for high-quality {PET} image synthesis}.
\newblock \bibinfo{journal}{Medical Image Analysis} \bibinfo{volume}{77}, \bibinfo{pages}{102335}.
\bibitem[{Maurer and Wang(2005)}]{2005Positron}
\bibinfo{author}{Maurer, L.}, \bibinfo{author}{Wang, J.}, \bibinfo{year}{2005}.
\newblock \bibinfo{title}{Positron {E}mission {T}omography: applications in drug discovery and drug development.}
\newblock \bibinfo{journal}{Current Topics in Medicinal Chemistry} \bibinfo{volume}{5}, \bibinfo{pages}{1053--1075}.
\bibitem[{Mehranian and Reader(2020)}]{mehranian2020model}
\bibinfo{author}{Mehranian, A.}, \bibinfo{author}{Reader, A.}, \bibinfo{year}{2020}.
\newblock \bibinfo{title}{Model-based deep learning {PET} image reconstruction using forward--backward splitting expectation--maximization}.
\newblock \bibinfo{journal}{IEEE Transactions on Radiation and Plasma Medical Sciences} \bibinfo{volume}{5}, \bibinfo{pages}{54--64}.
\bibitem[{Meyer et~al.(2020)Meyer, Cervenka, Kim, Kreisl, Henter and Innis}]{meyer2020neuroinflammation}
\bibinfo{author}{Meyer, J.H.}, \bibinfo{author}{Cervenka, S.}, \bibinfo{author}{Kim, M.}, \bibinfo{author}{Kreisl, W.C.}, \bibinfo{author}{Henter, I.D.}, \bibinfo{author}{Innis, R.B.}, \bibinfo{year}{2020}.
\newblock \bibinfo{title}{Neuroinflammation in psychiatric disorders: {PET} imaging and promising new targets}.
\newblock \bibinfo{journal}{The Lancet Psychiatry} .
\bibitem[{Mokri et~al.(2016)Mokri, Saripan, R., Nordin, Hashim and Marhaban}]{mokri2016pet}
\bibinfo{author}{Mokri, S.}, \bibinfo{author}{Saripan, M.}, \bibinfo{author}{R., A.}, \bibinfo{author}{Nordin, A.}, \bibinfo{author}{Hashim, S.}, \bibinfo{author}{Marhaban, M.}, \bibinfo{year}{2016}.
\newblock \bibinfo{title}{{PET} image reconstruction incorporating 3{D} mean-median sinogram filtering}.
\newblock \bibinfo{journal}{IEEE Transactions on Nuclear Science} \bibinfo{volume}{63}, \bibinfo{pages}{157--169}.
\bibitem[{Mosconi et~al.(2008)Mosconi, Tsui, Herholz, Pupi, Drzezga, Lucignani, Reiman, Holthoff, Kalbe and Sorbi}]{2008Multicenter}
\bibinfo{author}{Mosconi, L.}, \bibinfo{author}{Tsui, W.H.}, \bibinfo{author}{Herholz, K.}, \bibinfo{author}{Pupi, A.}, \bibinfo{author}{Drzezga, A.}, \bibinfo{author}{Lucignani, G.}, \bibinfo{author}{Reiman, E.M.}, \bibinfo{author}{Holthoff, V.}, \bibinfo{author}{Kalbe, E.}, \bibinfo{author}{Sorbi, S.}, \bibinfo{year}{2008}.
\newblock \bibinfo{title}{Multicenter standardized 18{F-FDG} {PET} diagnosis of mild cognitive impairment, {A}lzheimer\"s disease, and other dementias}.
\newblock \bibinfo{journal}{Journal of Nuclear Medicine} \bibinfo{volume}{49}, \bibinfo{pages}{390--398}.
\bibitem[{Newman(2013)}]{Newman2013}
\bibinfo{author}{Newman, M.E.J.}, \bibinfo{year}{2013}.
\newblock \bibinfo{title}{Network data}.
\newblock \bibinfo{howpublished}{\url{http://www-personal.umich.edu/~mejn/netdata/}}.
\bibitem[{Nguyen and Bai(2010)}]{nguyen2010cosine}
\bibinfo{author}{Nguyen, H.}, \bibinfo{author}{Bai, L.}, \bibinfo{year}{2010}.
\newblock \bibinfo{title}{Cosine similarity metric learning for face verification}.
\newblock \bibinfo{journal}{Asian Conference on Computer Vision} , \bibinfo{pages}{709--720}.
\bibitem[{Nguyen and Lee(2013)}]{nguyen2013incorporating}
\bibinfo{author}{Nguyen, V.}, \bibinfo{author}{Lee, S.}, \bibinfo{year}{2013}.
\newblock \bibinfo{title}{Incorporating anatomical side information into {PET} reconstruction using nonlocal regularization}.
\newblock \bibinfo{journal}{IEEE Transactions on Image Processing} \bibinfo{volume}{22}, \bibinfo{pages}{3961--3973}.
\bibitem[{Nichols et~al.(2002)Nichols, Qi, Asma and Leahy}]{nichols2002spatiotemporal}
\bibinfo{author}{Nichols, T.}, \bibinfo{author}{Qi, J.}, \bibinfo{author}{Asma, E.}, \bibinfo{author}{Leahy, R.}, \bibinfo{year}{2002}.
\newblock \bibinfo{title}{Spatiotemporal reconstruction of list-mode pet data}.
\newblock \bibinfo{journal}{IEEE Transactions on Medical Imaging} \bibinfo{volume}{21}, \bibinfo{pages}{396--404}.
\bibitem[{Olesen et~al.(2013)Olesen, Sullivan, Mulnix, Paulsen, Hojgaard, Roed, Carson, Morris and Larsen}]{2013List}
\bibinfo{author}{Olesen, O.V.}, \bibinfo{author}{Sullivan, J.M.}, \bibinfo{author}{Mulnix, T.}, \bibinfo{author}{Paulsen, R.R.}, \bibinfo{author}{Hojgaard, L.}, \bibinfo{author}{Roed, B.}, \bibinfo{author}{Carson, R.E.}, \bibinfo{author}{Morris, E.D.}, \bibinfo{author}{Larsen, R.}, \bibinfo{year}{2013}.
\newblock \bibinfo{title}{List-mode {PET} motion correction using markerless head tracking: {P}roof-of-concept with scans of human subject}.
\newblock \bibinfo{journal}{IEEE Transactions on Medical Imaging} \bibinfo{volume}{32}, \bibinfo{pages}{200--209}.
\bibitem[{Onishi et~al.(2021)Onishi, Hashimoto, Ote, Ohba, Ota, Yoshikawa and Ouchi}]{onishi2021anatomical}
\bibinfo{author}{Onishi, Y.}, \bibinfo{author}{Hashimoto, F.}, \bibinfo{author}{Ote, K.}, \bibinfo{author}{Ohba, H.}, \bibinfo{author}{Ota, R.}, \bibinfo{author}{Yoshikawa, E.}, \bibinfo{author}{Ouchi, Y.}, \bibinfo{year}{2021}.
\newblock \bibinfo{title}{Anatomical-guided attention enhances unsupervised {PET} image denoising performance}.
\newblock \bibinfo{journal}{Medical Image Analysis} \bibinfo{volume}{74}, \bibinfo{pages}{102226}.
\bibitem[{Ote et~al.(2020)Ote, Hashimoto, Kakimoto, Isobe, Inubushi, Ota, Tokui, Saito, Moriya and Omura}]{ote2020kinetics}
\bibinfo{author}{Ote, K.}, \bibinfo{author}{Hashimoto, F.}, \bibinfo{author}{Kakimoto, A.}, \bibinfo{author}{Isobe, T.}, \bibinfo{author}{Inubushi, T.}, \bibinfo{author}{Ota, R.}, \bibinfo{author}{Tokui, A.}, \bibinfo{author}{Saito, A.}, \bibinfo{author}{Moriya, T.}, \bibinfo{author}{Omura, T.}, \bibinfo{year}{2020}.
\newblock \bibinfo{title}{Kinetics-induced block matching and 5-{D} transform domain filtering for dynamic {PET} image denoising}.
\newblock \bibinfo{journal}{IEEE Transactions on Radiation and Plasma Medical Sciences} \bibinfo{volume}{4}, \bibinfo{pages}{720--728}.
\bibitem[{P. et~al.(2013)P., Hanzouli, Hatt, R. and Visvikis}]{le2013denoising}
\bibinfo{author}{P., L.}, \bibinfo{author}{Hanzouli, H.}, \bibinfo{author}{Hatt, M.}, \bibinfo{author}{R., L.}, \bibinfo{author}{Visvikis, D.}, \bibinfo{year}{2013}.
\newblock \bibinfo{title}{Denoising of {PET} images by combining wavelets and curvelets for improved preservation of resolution and quantitation}.
\newblock \bibinfo{journal}{Medical Image Analysis} \bibinfo{volume}{17}, \bibinfo{pages}{877--891}.
\bibitem[{Pan et~al.(2020)Pan, Liu, Lian, Xia and Shen}]{2020Spatially}
\bibinfo{author}{Pan, Y.}, \bibinfo{author}{Liu, M.}, \bibinfo{author}{Lian, C.}, \bibinfo{author}{Xia, Y.}, \bibinfo{author}{Shen, D.}, \bibinfo{year}{2020}.
\newblock \bibinfo{title}{Spatially-constrained fisher representation for brain disease identification with incomplete multi-modal neuroimages}.
\newblock \bibinfo{journal}{IEEE Transactions on Medical Imaging} \bibinfo{volume}{39}, \bibinfo{pages}{2965--2975}.
\bibitem[{Pan et~al.(2021)Pan, Liu, Xia and Shen}]{2021Disease}
\bibinfo{author}{Pan, Y.}, \bibinfo{author}{Liu, M.}, \bibinfo{author}{Xia, Y.}, \bibinfo{author}{Shen, D.}, \bibinfo{year}{2021}.
\newblock \bibinfo{title}{Disease-image-specific learning for diagnosis-oriented neuroimage synthesis with incomplete multi-modality data}.
\newblock \bibinfo{journal}{IEEE Transactions on Pattern Analysis and Machine Intelligence} \bibinfo{volume}{27}, \bibinfo{pages}{1675--1686}.
\bibitem[{Park et~al.(2019)Park, Liu, Wang and Zhu}]{park2019semantic}
\bibinfo{author}{Park, T.}, \bibinfo{author}{Liu, M.}, \bibinfo{author}{Wang, T.}, \bibinfo{author}{Zhu, J.}, \bibinfo{year}{2019}.
\newblock \bibinfo{title}{Semantic image synthesis with spatially-adaptive normalization}.
\newblock \bibinfo{journal}{Proceedings of the IEEE/CVF Conference on Computer Vision and Pattern Recognition} , \bibinfo{pages}{2337--2346}.
\bibitem[{Peiris et~al.(2022)Peiris, Hayat, Chen, Egan and Harandi}]{peiris2022robust}
\bibinfo{author}{Peiris, H.}, \bibinfo{author}{Hayat, M.}, \bibinfo{author}{Chen, Z.}, \bibinfo{author}{Egan, G.}, \bibinfo{author}{Harandi, M.}, \bibinfo{year}{2022}.
\newblock \bibinfo{title}{A robust volumetric transformer for accurate 3{D} tumor segmentation}.
\newblock \bibinfo{journal}{International Conference on Medical Image Computing and Computer-Assisted Intervention} , \bibinfo{pages}{162--172}.
\bibitem[{Ronneberger et~al.(2015)Ronneberger, Fischer and Brox}]{ronneberger2015u}
\bibinfo{author}{Ronneberger, O.}, \bibinfo{author}{Fischer, P.}, \bibinfo{author}{Brox, T.}, \bibinfo{year}{2015}.
\newblock \bibinfo{title}{U-net: {C}onvolutional networks for biomedical image segmentation}.
\newblock \bibinfo{journal}{International Conference on Medical image computing and computer-assisted intervention} , \bibinfo{pages}{234--241}.
\bibitem[{Sakthivel et~al.(2020)Sakthivel, Thakar, Prashanth, Angamuthu, Sharma and Kumar}]{sakthivel2020clinical}
\bibinfo{author}{Sakthivel, P.}, \bibinfo{author}{Thakar, A.}, \bibinfo{author}{Prashanth, A.}, \bibinfo{author}{Angamuthu, M.}, \bibinfo{author}{Sharma, S.C.}, \bibinfo{author}{Kumar, R.}, \bibinfo{year}{2020}.
\newblock \bibinfo{title}{Clinical applications of 68 ga-psma {PET/CT} on residual disease assessment of juvenile nasopharyngeal angiofibroma ({JNA})}.
\newblock \bibinfo{journal}{Nuclear Medicine and Molecular Imaging} \bibinfo{volume}{54}, \bibinfo{pages}{63--64}.
\bibitem[{Schaefferkoetter et~al.(2020)Schaefferkoetter, Yan, Ortega, Sertic, Lechtman, Eshet, Metser and Veit-Haibach}]{schaefferkoetter2020convolutional}
\bibinfo{author}{Schaefferkoetter, J.}, \bibinfo{author}{Yan, J.}, \bibinfo{author}{Ortega, C.}, \bibinfo{author}{Sertic, A.}, \bibinfo{author}{Lechtman, E.}, \bibinfo{author}{Eshet, Y.}, \bibinfo{author}{Metser, U.}, \bibinfo{author}{Veit-Haibach, P.}, \bibinfo{year}{2020}.
\newblock \bibinfo{title}{Convolutional neural networks for improving image quality with noisy pet data}.
\newblock \bibinfo{journal}{EJNMMI research} \bibinfo{volume}{10}, \bibinfo{pages}{1--11}.
\bibitem[{Shukla(2003)}]{shukla2003complex}
\bibinfo{author}{Shukla, P.D.}, \bibinfo{year}{2003}.
\newblock \bibinfo{title}{Complex wavelet transforms and their applications}.
\newblock \bibinfo{journal}{A Dissertation Submitted of Signal Processing Division, Department of Electronic and Electrical Engineering University of Strathclyde Scotland United Kingdom} .
\bibitem[{Slovis and Thomas(2002)}]{Slovis2002The}
\bibinfo{author}{Slovis}, \bibinfo{author}{Thomas, L.}, \bibinfo{year}{2002}.
\newblock \bibinfo{title}{The {ALARA} concept in pediatric {CT}: myth or reality?}
\newblock \bibinfo{journal}{Radiology} \bibinfo{volume}{223}, \bibinfo{pages}{5--6}.
\bibitem[{Song et~al.(2021)Song, Yang and Dutta}]{song2021noise2void}
\bibinfo{author}{Song, T.}, \bibinfo{author}{Yang, F.}, \bibinfo{author}{Dutta, J.}, \bibinfo{year}{2021}.
\newblock \bibinfo{title}{Noise2{V}oid: unsupervised denoising of {PET} images}.
\newblock \bibinfo{journal}{Physics in Medicine \& Biology} \bibinfo{volume}{66}, \bibinfo{pages}{214002}.
\bibitem[{Souza et~al.(2019)Souza, Lebel and Frayne}]{souza2019hybrid}
\bibinfo{author}{Souza, R.}, \bibinfo{author}{Lebel, R.M.}, \bibinfo{author}{Frayne, R.}, \bibinfo{year}{2019}.
\newblock \bibinfo{title}{A hybrid, dual domain, cascade of convolutional neural networks for magnetic resonance image reconstruction}.
\newblock \bibinfo{journal}{International Conference on Medical Imaging with Deep Learning} , \bibinfo{pages}{437--446}.
\bibitem[{Spencer et~al.(2021)Spencer, Berg, Schmall, Omidvari, Leung, Abdelhafez, Tang, Deng, Dong and Lv}]{spencer2021performance}
\bibinfo{author}{Spencer, B.}, \bibinfo{author}{Berg, E.}, \bibinfo{author}{Schmall, J.}, \bibinfo{author}{Omidvari, N.}, \bibinfo{author}{Leung, E.}, \bibinfo{author}{Abdelhafez, Y.G.}, \bibinfo{author}{Tang, S.}, \bibinfo{author}{Deng, Z.}, \bibinfo{author}{Dong, Y.}, \bibinfo{author}{Lv, Y.}, \bibinfo{year}{2021}.
\newblock \bibinfo{title}{Performance evaluation of the u{EXPLORER} total-body {PET/CT} scanner based on {NEMA NU} 2-2018 with additional tests to characterize pet scanners with a long axial field of view}.
\newblock \bibinfo{journal}{Journal of Nuclear Medicine} \bibinfo{volume}{62}, \bibinfo{pages}{861--870}.
\bibitem[{Spurr et~al.(2017)Spurr, Aksan and Hilliges}]{spurr2017guiding}
\bibinfo{author}{Spurr, A.}, \bibinfo{author}{Aksan, E.}, \bibinfo{author}{Hilliges, O.}, \bibinfo{year}{2017}.
\newblock \bibinfo{title}{Guiding info{GAN} with semi-supervision}.
\newblock \bibinfo{journal}{Joint European Conference on Machine Learning and Knowledge Discovery in Databases} , \bibinfo{pages}{119--134}.
\bibitem[{Sun et~al.(2020)Sun, Wang, Huang, Ding, Greenspan and Paisley}]{sun2020adversarial}
\bibinfo{author}{Sun, L.}, \bibinfo{author}{Wang, J.}, \bibinfo{author}{Huang, Y.}, \bibinfo{author}{Ding, X.}, \bibinfo{author}{Greenspan, H.}, \bibinfo{author}{Paisley, J.}, \bibinfo{year}{2020}.
\newblock \bibinfo{title}{An adversarial learning approach to medical image synthesis for lesion detection}.
\newblock \bibinfo{journal}{IEEE Journal of Biomedical and Health Informatics} \bibinfo{volume}{24}, \bibinfo{pages}{2303--2314}.
\bibitem[{Tang and Rahmim(2014)}]{tang2014anatomy}
\bibinfo{author}{Tang, J.}, \bibinfo{author}{Rahmim, A.}, \bibinfo{year}{2014}.
\newblock \bibinfo{title}{Anatomy assisted {PET} image reconstruction incorporating multi-resolution joint entropy}.
\newblock \bibinfo{journal}{Physics in Medicine \& Biology} \bibinfo{volume}{60}, \bibinfo{pages}{31}.
\bibitem[{Toft(1996)}]{toft1996radon}
\bibinfo{author}{Toft, P.}, \bibinfo{year}{1996}.
\newblock \bibinfo{title}{The radon transform}.
\newblock \bibinfo{journal}{Theory and Implementation (Ph. D. Dissertation)(Copenhagen: Technical University of Denmark)} .
\bibitem[{Ulyanov et~al.(2016)Ulyanov, Vedaldi and Lempitsky}]{ulyanov2016instance}
\bibinfo{author}{Ulyanov, D.}, \bibinfo{author}{Vedaldi, A.}, \bibinfo{author}{Lempitsky, V.}, \bibinfo{year}{2016}.
\newblock \bibinfo{title}{Instance normalization: {T}he missing ingredient for fast stylization}.
\newblock \bibinfo{journal}{arXiv preprint arXiv:1607.08022} .
\bibitem[{Umehara et~al.(2017)Umehara, Ota and Ishida}]{2017Super}
\bibinfo{author}{Umehara, K.}, \bibinfo{author}{Ota, J.}, \bibinfo{author}{Ishida, T.}, \bibinfo{year}{2017}.
\newblock \bibinfo{title}{Super-resolution imaging of mammograms based on the super-resolution convolutional neural network}.
\newblock \bibinfo{journal}{Open Journal of Medical Imaging} \bibinfo{volume}{07}, \bibinfo{pages}{180--195}.
\bibitem[{Wang et~al.(2016)Wang, Ma, An, Shi, Zhang, Wu, Zhou and Shen}]{2016Semisupervised}
\bibinfo{author}{Wang, Y.}, \bibinfo{author}{Ma, G.}, \bibinfo{author}{An, L.}, \bibinfo{author}{Shi, F.}, \bibinfo{author}{Zhang, P.}, \bibinfo{author}{Wu, X.}, \bibinfo{author}{Zhou, J.}, \bibinfo{author}{Shen, D.}, \bibinfo{year}{2016}.
\newblock \bibinfo{title}{Semi-supervised tripled dictionary learning for standard-dose {PET} image prediction using low-dose {PET} and multimodal {MRI}}.
\newblock \bibinfo{journal}{IEEE Transactions on Biomedical Engineering} \bibinfo{volume}{64}, \bibinfo{pages}{569--579}.
\bibitem[{Wang et~al.(2018)Wang, Yu, Wang, Zu, Lalush, Lin, Wu, Zhou, Shen and Zhou}]{20183D}
\bibinfo{author}{Wang, Y.}, \bibinfo{author}{Yu, B.}, \bibinfo{author}{Wang, L.}, \bibinfo{author}{Zu, C.}, \bibinfo{author}{Lalush, D.S.}, \bibinfo{author}{Lin, W.}, \bibinfo{author}{Wu, X.}, \bibinfo{author}{Zhou, J.}, \bibinfo{author}{Shen, D.}, \bibinfo{author}{Zhou, L.}, \bibinfo{year}{2018}.
\newblock \bibinfo{title}{3{D} conditional generative adversarial networks for high-quality {PET} image estimation at low dose}.
\newblock \bibinfo{journal}{Neuroimage} \bibinfo{volume}{174}, \bibinfo{pages}{550--562}.
\bibitem[{Wang et~al.(2015)Wang, Zhang, An, Ma, Kang, Shi, Wu, Zhou, Lalush and Lin}]{2015Predicting}
\bibinfo{author}{Wang, Y.}, \bibinfo{author}{Zhang, P.}, \bibinfo{author}{An, L.}, \bibinfo{author}{Ma, G.}, \bibinfo{author}{Kang, J.}, \bibinfo{author}{Shi, F.}, \bibinfo{author}{Wu, X.}, \bibinfo{author}{Zhou, J.}, \bibinfo{author}{Lalush, D.S.}, \bibinfo{author}{Lin, W.}, \bibinfo{year}{2015}.
\newblock \bibinfo{title}{Predicting standard-dose {PET} image from low-dose {PET} and multimodal {MR} images using mapping-based sparse representation}.
\newblock \bibinfo{journal}{Physics in Medicine \& Biology} \bibinfo{volume}{61}, \bibinfo{pages}{791--801}.
\bibitem[{Wang et~al.(2019)Wang, Zhou, Yu, Wang, Zu, Lalush, Lin, Wu, Zhou and Shen}]{20193D}
\bibinfo{author}{Wang, Y.}, \bibinfo{author}{Zhou, L.}, \bibinfo{author}{Yu, B.}, \bibinfo{author}{Wang, L.}, \bibinfo{author}{Zu, C.}, \bibinfo{author}{Lalush, D.S.}, \bibinfo{author}{Lin, W.}, \bibinfo{author}{Wu, X.}, \bibinfo{author}{Zhou, J.}, \bibinfo{author}{Shen, D.}, \bibinfo{year}{2019}.
\newblock \bibinfo{title}{3{D} auto-context-based locality adaptive multi-modality {GAN}s for {PET} synthesis}.
\newblock \bibinfo{journal}{IEEE Transactions on Medical Imaging} \bibinfo{volume}{38}, \bibinfo{pages}{1328--1339}.
\bibitem[{Wang et~al.(2004)Wang, Bovik, Sheikh and Simoncelli}]{wang2004image}
\bibinfo{author}{Wang, Z.}, \bibinfo{author}{Bovik, A.}, \bibinfo{author}{Sheikh, H.}, \bibinfo{author}{Simoncelli, E.}, \bibinfo{year}{2004}.
\newblock \bibinfo{title}{Image quality assessment: from error visibility to structural similarity}.
\newblock \bibinfo{journal}{IEEE Transactions on Image Processing} \bibinfo{volume}{13}, \bibinfo{pages}{600--612}.
\bibitem[{Wasserthal et~al.(2022)Wasserthal, Meyer, Breit, Cyriac, Yang and Segeroth}]{wasserthal2022totalsegmentator}
\bibinfo{author}{Wasserthal, J.}, \bibinfo{author}{Meyer, M.}, \bibinfo{author}{Breit, H.}, \bibinfo{author}{Cyriac, J.}, \bibinfo{author}{Yang, S.}, \bibinfo{author}{Segeroth, M.}, \bibinfo{year}{2022}.
\newblock \bibinfo{title}{Totalsegmentator: robust segmentation of 104 anatomical structures in {CT} images}.
\newblock \bibinfo{journal}{arXiv preprint arXiv:2208.05868} .
\bibitem[{Williams et~al.(2005)Williams, Blake and Cipolla}]{2005Sparse}
\bibinfo{author}{Williams, O.}, \bibinfo{author}{Blake, A.}, \bibinfo{author}{Cipolla, R.}, \bibinfo{year}{2005}.
\newblock \bibinfo{title}{Sparse {B}ayesian learning for efficient visual tracking}.
\newblock \bibinfo{journal}{IEEE Transactions on Pattern Analysis and Machine Intelligence} \bibinfo{volume}{27}, \bibinfo{pages}{1292--1304}.
\bibitem[{Willmott and Matsuura(2005)}]{willmott2005advantages}
\bibinfo{author}{Willmott, C.}, \bibinfo{author}{Matsuura, K.}, \bibinfo{year}{2005}.
\newblock \bibinfo{title}{Advantages of the mean absolute error ({MAE}) over the root mean square error ({RMSE}) in assessing average model performance}.
\newblock \bibinfo{journal}{Climate Research} \bibinfo{volume}{30}, \bibinfo{pages}{79--82}.
\bibitem[{Wu et~al.(2011)Wu, Jia, Wang and Shen}]{wu2011sharpmean}
\bibinfo{author}{Wu, G.}, \bibinfo{author}{Jia, H.}, \bibinfo{author}{Wang, Q.}, \bibinfo{author}{Shen, D.}, \bibinfo{year}{2011}.
\newblock \bibinfo{title}{Sharp{M}ean: groupwise registration guided by sharp mean image and tree-based registration}.
\newblock \bibinfo{journal}{NeuroImage} \bibinfo{volume}{56}, \bibinfo{pages}{1968--1981}.
\bibitem[{Wu et~al.(2021)Wu, Hu, Niu, Yu, Vardhanabhuti and Wang}]{wu2021drone}
\bibinfo{author}{Wu, W.}, \bibinfo{author}{Hu, D.}, \bibinfo{author}{Niu, C.}, \bibinfo{author}{Yu, H.}, \bibinfo{author}{Vardhanabhuti, V.}, \bibinfo{author}{Wang, G.}, \bibinfo{year}{2021}.
\newblock \bibinfo{title}{D{RONE}: Dual-domain residual-based optimization network for sparse-view {CT} reconstruction}.
\newblock \bibinfo{journal}{IEEE Transactions on Medical Imaging} .
\bibitem[{Wu and He(2018)}]{wu2018group}
\bibinfo{author}{Wu, Y.}, \bibinfo{author}{He, K.}, \bibinfo{year}{2018}.
\newblock \bibinfo{title}{Group normalization}.
\newblock \bibinfo{journal}{Proceedings of the European Conference on Computer Vsion (ECCV)} , \bibinfo{pages}{3--19}.
\bibitem[{Xiang et~al.(2017)Xiang, Qiao, Nie, An, Lin, Wang and Shen}]{2017Deep}
\bibinfo{author}{Xiang, L.}, \bibinfo{author}{Qiao, Y.}, \bibinfo{author}{Nie, D.}, \bibinfo{author}{An, L.}, \bibinfo{author}{Lin, W.}, \bibinfo{author}{Wang, Q.}, \bibinfo{author}{Shen, D.}, \bibinfo{year}{2017}.
\newblock \bibinfo{title}{Deep auto-context convolutional neural networks for standard-dose {PET} image estimation from low-dose {PET/MRI}}.
\newblock \bibinfo{journal}{Neurocomputing} \bibinfo{volume}{267}, \bibinfo{pages}{406--416}.
\bibitem[{Xiang et~al.(2020)Xiang, Wang, Gong, Zaharchuk and Zhang}]{2020Noise}
\bibinfo{author}{Xiang, L.}, \bibinfo{author}{Wang, L.}, \bibinfo{author}{Gong, E.}, \bibinfo{author}{Zaharchuk, G.}, \bibinfo{author}{Zhang, T.}, \bibinfo{year}{2020}.
\newblock \bibinfo{title}{Noise-aware standard-dose {PET} reconstruction using general and adaptive robust loss}.
\newblock \bibinfo{journal}{International Workshop on Machine Learning in Medical Imaging} , \bibinfo{pages}{654--662}.
\bibitem[{Xiao et~al.(2021)Xiao, Yu, Xing, Yuille and Zhou}]{xiao2021dualnorm}
\bibinfo{author}{Xiao, J.}, \bibinfo{author}{Yu, L.}, \bibinfo{author}{Xing, L.}, \bibinfo{author}{Yuille, A.}, \bibinfo{author}{Zhou, Y.}, \bibinfo{year}{2021}.
\newblock \bibinfo{title}{Dual{N}orm-{UN}et: {I}ncorporating global and local statistics for robust medical image segmentation}.
\newblock \bibinfo{journal}{arXiv preprint arXiv:2103.15858} .
\bibitem[{Xu et~al.(2017)Xu, Gong, Pauly and Zaharchuk}]{xu2017200x}
\bibinfo{author}{Xu, J.}, \bibinfo{author}{Gong, E.}, \bibinfo{author}{Pauly, J.}, \bibinfo{author}{Zaharchuk, G.}, \bibinfo{year}{2017}.
\newblock \bibinfo{title}{200x low-dose {PET} reconstruction using deep learning}.
\newblock \bibinfo{journal}{arXiv preprint arXiv:1712.04119} .
\bibitem[{Xuan et~al.(2020)Xuan, Si, Zhang, Xue and Wang}]{2020Reduce}
\bibinfo{author}{Xuan, K.}, \bibinfo{author}{Si, L.}, \bibinfo{author}{Zhang, L.}, \bibinfo{author}{Xue, Z.}, \bibinfo{author}{Wang, Q.}, \bibinfo{year}{2020}.
\newblock \bibinfo{title}{Reduce slice spacing of {MR} images by super-resolution learned without ground-truth}.
\newblock \bibinfo{journal}{arXiv preprint arXiv:2003.12627} .
\bibitem[{Yan et~al.(2015)Yan, Lim and Townsend}]{yan2015mri}
\bibinfo{author}{Yan, J.}, \bibinfo{author}{Lim, J.}, \bibinfo{author}{Townsend, D.}, \bibinfo{year}{2015}.
\newblock \bibinfo{title}{M{RI}-guided brain {PET} image filtering and partial volume correction}.
\newblock \bibinfo{journal}{Physics in Medicine \& Biology} \bibinfo{volume}{60}, \bibinfo{pages}{961}.
\bibitem[{Yang et~al.(2023)Yang, He and Zhang}]{yang2023alternating}
\bibinfo{author}{Yang, D.}, \bibinfo{author}{He, X.}, \bibinfo{author}{Zhang, R.}, \bibinfo{year}{2023}.
\newblock \bibinfo{title}{Alternating attention {T}ransformer for single image deraining}.
\newblock \bibinfo{journal}{Digital Signal Processing} , \bibinfo{pages}{104144}.
\bibitem[{Yang et~al.(2020)Yang, Sun, Carass, Zhao, Lee, Prince and Xu}]{yang2020unsupervised}
\bibinfo{author}{Yang, H.}, \bibinfo{author}{Sun, J.}, \bibinfo{author}{Carass, A.}, \bibinfo{author}{Zhao, C.}, \bibinfo{author}{Lee, J.}, \bibinfo{author}{Prince, J.}, \bibinfo{author}{Xu, Z.}, \bibinfo{year}{2020}.
\newblock \bibinfo{title}{Unsupervised {MR}-to-{CT} synthesis using structure-constrained cycle{GAN}}.
\newblock \bibinfo{journal}{IEEE Transactions on Medical Imaging} \bibinfo{volume}{39}, \bibinfo{pages}{4249--4261}.
\bibitem[{Yi et~al.(2019)Yi, Walia and Babyn}]{yi2019generative}
\bibinfo{author}{Yi, X.}, \bibinfo{author}{Walia, E.}, \bibinfo{author}{Babyn, P.}, \bibinfo{year}{2019}.
\newblock \bibinfo{title}{Generative adversarial network in medical imaging: A review}.
\newblock \bibinfo{journal}{Medical image analysis} \bibinfo{volume}{58}, \bibinfo{pages}{101552}.
\bibitem[{Yie et~al.(2020)Yie, Kang, Hwang and Lee}]{yie2020self}
\bibinfo{author}{Yie, S.}, \bibinfo{author}{Kang, S.}, \bibinfo{author}{Hwang, D.}, \bibinfo{author}{Lee, J.}, \bibinfo{year}{2020}.
\newblock \bibinfo{title}{Self-supervised {PET} denoising}.
\newblock \bibinfo{journal}{Nuclear Medicine and Molecular Imaging} \bibinfo{volume}{54}, \bibinfo{pages}{299--304}.
\bibitem[{Yurt et~al.(2021)Yurt, Dar, Erdem, Erdem, Oguz and {\c{C}}ukur}]{yurt2021mustgan}
\bibinfo{author}{Yurt, M.}, \bibinfo{author}{Dar, S.}, \bibinfo{author}{Erdem, A.}, \bibinfo{author}{Erdem, E.}, \bibinfo{author}{Oguz, K.}, \bibinfo{author}{{\c{C}}ukur, T.}, \bibinfo{year}{2021}.
\newblock \bibinfo{title}{must{GAN}: multi-stream generative adversarial networks for {MR} image synthesis}.
\newblock \bibinfo{journal}{Medical Image Analysis} \bibinfo{volume}{70}, \bibinfo{pages}{101--144}.
\bibitem[{Zeng et~al.(2023)Zeng, Gao, Hu, Feng, Hou, Rong and Wang}]{zeng2023ss}
\bibinfo{author}{Zeng, L.L.}, \bibinfo{author}{Gao, K.}, \bibinfo{author}{Hu, D.}, \bibinfo{author}{Feng, Z.}, \bibinfo{author}{Hou, C.}, \bibinfo{author}{Rong, P.}, \bibinfo{author}{Wang, W.}, \bibinfo{year}{2023}.
\newblock \bibinfo{title}{S{S-TBN}: a semi-supervised tri-branch network for {COVID-19} screening and lesion segmentation}.
\newblock \bibinfo{journal}{IEEE Transactions on Pattern Analysis and Machine Intelligence} .
\bibitem[{Zhang et~al.(2017)Zhang, Zuo, Chen, Meng and Zhang}]{zhang2017beyond}
\bibinfo{author}{Zhang, K.}, \bibinfo{author}{Zuo, W.}, \bibinfo{author}{Chen, Y.}, \bibinfo{author}{Meng, D.}, \bibinfo{author}{Zhang, L.}, \bibinfo{year}{2017}.
\newblock \bibinfo{title}{Beyond a gaussian denoiser: Residual learning of deep {CNN} for image denoising}.
\newblock \bibinfo{journal}{IEEE Transactions on Image Processing} \bibinfo{volume}{26}, \bibinfo{pages}{3142--3155}.
\bibitem[{Zhang et~al.(2020)Zhang, Xie, Berg, Judenhofer, Liu, Xu, Ding, Lv, Dong and Deng}]{zhang2020total}
\bibinfo{author}{Zhang, X.}, \bibinfo{author}{Xie, Z.}, \bibinfo{author}{Berg, E.}, \bibinfo{author}{Judenhofer, M.S.}, \bibinfo{author}{Liu, W.}, \bibinfo{author}{Xu, T.}, \bibinfo{author}{Ding, Y.}, \bibinfo{author}{Lv, Y.}, \bibinfo{author}{Dong, Y.}, \bibinfo{author}{Deng, Z.}, \bibinfo{year}{2020}.
\newblock \bibinfo{title}{Total-body dynamic reconstruction and parametric imaging on the u{EXPLORER}}.
\newblock \bibinfo{journal}{Journal of Nuclear Medicine} \bibinfo{volume}{61}, \bibinfo{pages}{285--291}.
\bibitem[{Zhou et~al.(2022)Zhou, Chen, Zhou, Duncan and Liu}]{zhou2022dudodr}
\bibinfo{author}{Zhou, B.}, \bibinfo{author}{Chen, X.}, \bibinfo{author}{Zhou, S.}, \bibinfo{author}{Duncan, J.}, \bibinfo{author}{Liu, C.}, \bibinfo{year}{2022}.
\newblock \bibinfo{title}{Du{DoDR}-{Net}: Dual-domain data consistent recurrent network for simultaneous sparse view and metal artifact reduction in computed tomography}.
\newblock \bibinfo{journal}{Medical Image Analysis} \bibinfo{volume}{75}, \bibinfo{pages}{102289}.
\bibitem[{Zhou et~al.(2020)Zhou, Schaefferkoetter, Tham, Huang and Yan}]{zhou2020supervised}
\bibinfo{author}{Zhou, L.}, \bibinfo{author}{Schaefferkoetter, J.}, \bibinfo{author}{Tham, I.}, \bibinfo{author}{Huang, G.}, \bibinfo{author}{Yan, J.}, \bibinfo{year}{2020}.
\newblock \bibinfo{title}{Supervised learning with {C}ycle{GAN} for low-dose {FDG} {PET} image denoising}.
\newblock \bibinfo{journal}{Medical Image Analysis} \bibinfo{volume}{65}, \bibinfo{pages}{101770}.
\bibitem[{Zhu et~al.(2017)Zhu, Park, Isola and Efros}]{2017Unpaired}
\bibinfo{author}{Zhu, J.Y.}, \bibinfo{author}{Park, T.}, \bibinfo{author}{Isola, P.}, \bibinfo{author}{Efros, A.A.}, \bibinfo{year}{2017}.
\newblock \bibinfo{title}{Unpaired image-to-image translation using cycle-consistent adversarial networks}.
\newblock \bibinfo{journal}{Proceedings of the IEEE International Conference on Computer Vision} , \bibinfo{pages}{2223--2232}.

\end{thebibliography}


\end{document}